\documentclass[12pt]{article}
\usepackage{epsfig,amssymb}

\newcommand{\bq}{\begin{equation}}
\newcommand{\eq}{\end{equation}}
\newcommand{\bqa}{\begin{eqnarray}}
\newcommand{\eqa}{\end{eqnarray}}
\newcommand{\ben}{\begin{enumerate}}
\newcommand{\een}{\end{enumerate}}
\newcommand{\bc}{\begin{center}}
\newcommand{\ec}{\end{center}}
\newcommand{\bqb}{\begin{eqnarray*}}
\newcommand{\eqb}{\end{eqnarray*}}

\def\gsim{\gtrsim}
\def\lsim{\lesssim}

\def\vectrl #1{\buildrel\leftrightarrow \over #1}
\def\partrl{\vectrl{\partial}}

\def\ie{{\it i.e. }}
\def\eg{{\it e.g. }}

\def\etal{{\it et.al. }}

\def\L{ {\cal L }}
\def\H{ {\cal H }}
\def\F{ {\cal F }}

\def\N{ {\cal N }}

\def\B{\tilde {\cal B}}
\def\Del{\tilde { \Delta}}
\def\sw{s_W}
\def\cw{c_W}
\def\swd{s^2_W}
\def\cwd{c^2_W}

\def\mwd{m_W^2}
\def\mw{m_W}
\def\mz{m_Z}
\def\mzd{m_Z^2}

\def\t{\hat t}
\def\s{\hat s}
\def\u{\hat u}
\def\Egg{\sqrt{s_{\gamma \gamma}}}
\def\rtau{\sqrt{\tau}}

\def\Sn#1{\mathrm{Sign} #1 }

\def\tchi{\tilde \chi}
\def\Stop{\tilde t}
\def\cfR{\cos\phi_R}
\def\cfL{\cos\phi_L}
\def\sfR{\sin\phi_R}
\def\sfL{\sin\phi_L}
\def\calpha{\cos \alpha}
\def\cbeta{\cos\beta}
\def\salpha{\sin \alpha}
\def\sbeta{\sin \beta}

\def\pr#1#2#3{ Phys. Rev. ${\bf{#1}}$, #2 (#3)}

\def\pl#1#2#3{ Phys. Lett. ${\bf{#1}}$, #2 (#3)}
\def\prep#1#2#3{ Phys. Rep. ${\bf{#1}}$, #2 (#3)}

\def\np#1#2#3{ Nucl. Phys. ${\bf{#1}}$, #2 (#3)}
\def\zp#1#2#3{ Z. f. Phys. ${\bf{#1}}$, #2 (#3)}
\def\epj#1#2#3{ Eur. Phys. J. ${\bf{#1}}$, #2 (#3)}

\def\cpc#1#2#3{Comput. Phys. Commun. ${\bf{#1}}$, #2 (#3)}

\def\AnnPhys#1#2#3{Ann. Phys. (N.Y.)${\bf{#1}}$, #2 (#3)}
\def\prog#1#2#3{ Prog. Theor. Phys. ${\bf{#1}}$, #2 (#3)}

\begin{document}
\pagenumbering{arabic}
\thispagestyle{empty}
\def\thefootnote{\fnsymbol{footnote}}
\setcounter{footnote}{1}

\begin{flushright}
PM/00-27 \\
THES-TP 2000/08 \\
hep-ph/0010006 \\
September 2000\\
Improved version
 \end{flushright}
\vspace{2cm}
\begin{center}
{\Large\bf The heavy neutral Higgs signature in
the $\gamma \gamma \to Z Z $ process.}\footnote{Partially
supported by the European
Community grant  HPRN-CT-2000-00149.}
 \vspace{1.5cm}  \\
{\large G.J. Gounaris$^a$, P.I. Porfyriadis$^a$ and
F.M. Renard$^b$}\\
\vspace{0.7cm}
$^a$Department of Theoretical Physics, Aristotle
University of Thessaloniki,\\
Gr-54006, Thessaloniki, Greece.\\
\vspace{0.2cm}
$^b$Physique
Math\'{e}matique et Th\'{e}orique,
UMR 5825\\
Universit\'{e} Montpellier II,
 F-34095 Montpellier Cedex 5.\\
\vspace{0.2cm}

 \vspace*{1cm}

{\bf Abstract}
\end{center}

If the Standard Model (SM) Higgs particle
is sufficiently heavy, then its  contribution to
$\gamma \gamma \to Z Z$
should be largely imaginary,   interfering  with the  also
predominantly imaginary SM "background"generated
by the $W$-loop. For standard model
 Higgs masses in the region $200\lsim m_H \lsim 500$ GeV,
this interference is found to be  constructive and
increasing  the Higgs signal.
 In the minimal SUSY case an interference   effect
  should also  appear for  the contribution of the  heavier
 CP-even neutral Higgs boson  $H^0$, provided it is sufficiently
heavy. The effect  is somewhat reduced though,
by the smallness of the $H^0$  width and the $\gamma\gamma$ and  $ZZ$
branching ratios.  The interference is again found to be
 constructive
 for part of the parameter space corresponding to   sfermion
 masses  at the TeV scale and maximal  stop mixing.
For both the SM and SUSY cases, regions of the parameter space
exist though, where the interference may be destructive.
It is therefore essential to take these effects into account
when searching for possible scalar
Higgs-like candidates. To this aim, we present the complete
analytic expressions for both resonance and background
amplitudes.

\def\thefootnote{\arabic{footnote}}
\setcounter{footnote}{0}
\clearpage

\section{Introduction}

Searching for the Higgs particle(s) is definitely the central
 aim in particle
physics at present.  If the Standard Model (SM)  correctly describes
nature, then the present LEP results require the  Higgs mass
to be  heavier than 113 GeV \cite{Higgs-exp}.
This constraint is somewhat loosened in  minimal SUSY,
in which for
typical scenarios assuming sfermion masses at the TeV scale
and maximal stop mixing,  the  lower bound  on the mass of the
lightest CP-even neutral Higgs $h^0$ is reduced to about 90 GeV,
while  the  $\tan\beta$-region $(0.5 ~-~ 2.3)$ is excluded
\cite{Higgs-exp}.

After the discovery of the Higgs particle(s), the necessity will of
course arise to  secure
its identification. To this aim, a  photon-photon  Collider
($LC_{\gamma \gamma}$) realized  through the
laser backscattering method \cite{LCgg}  in a high luminosity
$e^-e^-$ or $e^+e^-$ Collider  (LC) \cite{LC}, should be very
useful. In an $LC_{\gamma \gamma}$ the neutral Higgs
particle may then be produced directly in the $s$-channel,
and if it is not too narrow,  even its line shape may be studied.
\par

For a standard model (SM) light Higgs boson
(i.e. $m_H \lsim 135 ~ \rm GeV$)
the rate of direct production in $\gamma\gamma\to H$ is indeed
very high and the detection of the Higgs boson should clearly be
done through the dominant decay channel $H\to b\bar b$
\cite{revggH, Melles, Jikia-Soldner}.
For higher Higgs masses though, the situation changes because the Higgs
becomes broader and  the dominant channels are now
$WW$ and $ZZ$. A very  interesting channel
for Higgs detection is then  the  $ZZ$ one, in which at least one $Z$
decays into lepton pairs and the other one  into hadrons or
leptons. Such a channel will be very interesting, even though
$Br(H \to \gamma \gamma)$ decreases rapidly as the
Higgs mass increases.\par

However, to this $\gamma \gamma \to H \to ZZ$ channel,
there is an important
$\gamma\gamma\to ZZ$ background process
arising mainly through  $W$ and fermions box-type
contributions \cite{Glover-ZZ, Jikia-ZZ, ggZZ}.
As it has  already been noticed in \cite{Berger-Chanowitz}  from
the study of strong $WW$ interactions,  it is possible
to enhance the signal relative to the background by using polarized
photon beams and applying suitable cuts on the decay products
of the $Z$ bosons.

But, as emphasized in \cite{ggZZ}, this
background process has the  remarkable property that
 at high energies its predominant  helicity amplitudes
are almost purely imaginary and conserve
helicity\footnote{A similar property has also been observed
for  the processes $\gamma\gamma\to \gamma\gamma,~\gamma Z$
at sufficiently  high energies \cite{ gggg, gggZ}.}.
Thus, important  interference
effects between the Higgs and  background contributions
may appear, which should be taken into account
when analyzing  experimental data.\par

 The first aim of this paper is to explore  this interference
phenomenon in SM  for Higgs masses above the ZZ production threshold,
using the already known one-loop $\gamma\gamma\to ZZ$ amplitudes
\cite{Glover-ZZ, Jikia-ZZ, ggZZ}. \par

We next turn to  a general MSSM model \cite{SUSY-reviews},
assuming  no CP violation  other than the standard one
contained  in the   Yukawa sector. In this case,
the Higgs boson spectrum is much richer, with
two CP-even scalars $h^0$, $H^0$ and one CP-odd pseudoscalar $A^0$
\cite{Gunion-book}. We consider SUSY scenarios in which
$H^0$ and $A^0$ are  heavier than about 200 ~GeV, while
the SUSY breaking sfermion parameters are taken at the TeV scale
and  the stop mixing is maximal. Such scenarios have the tendency
to lead to an $h^0$ which is well within the presently
experimentally allowed region \cite{Higgs-exp}.
The lightest Higgs boson vertex
$h^0\gamma \gamma $ for SUSY models, has recently been
studied in \cite{h0gg-Djouadi}.
Since the CP-odd $A^0$ has no tree-level
coupling to $\gamma\gamma$ or  $ZZ$,  the
 $\gamma \gamma \to ZZ$  channel may be used
for the on-shell production of  the CP-even Higgs $H^0$.\par

Apart from the existence of the lighter Higgs $h^0$, there are  several
new features discriminating  the heavier  SUSY $H^0$ boson,
 from the case of a heavy standard Higgs.
The decay spectrum of the SUSY $H^0$ is expected to differ
from that of a
heavy standard Higgs, because of the possible appearance
of new decay channels and mixing effects which strongly influence
its couplings  to gauge bosons.
Thus for $m_{H^0} \gsim 200 \rm GeV$,  the SUSY $H^0$
is expected to be much narrower than a heavy standard $H$,
 and its branching ratios
$Br(H^0 \to\gamma\gamma)$ and $Br(H^0 \to ZZ)$ much smaller
\cite{Gunion-book, single-H0A0}. Moreover,
in an MSSM description, the $\gamma\gamma\to ZZ$ background
 receives at one loop new contributions from virtual
SUSY partners running inside the loop \cite{ggZZ}.
So finally, within the MSSM,
the treatment of the Higgs effects in the $\gamma\gamma\to ZZ$
process requires a specific analysis. This constitutes the
second topic of this paper.\par

In Section 2 we write the Higgs contributions to the
$\gamma\gamma\to ZZ$ amplitudes for the SM ($H$)
and for the MSSM ($h^0,~H^0$) cases. We give the explicit
expressions of the one loop Higgs couplings to $\gamma\gamma$ and
the tree level couplings to $ZZ$. In the Appendix  we collect
all background amplitudes for $\gamma\gamma\to ZZ$  in SM
and in MSSM. They are taken from  \cite{ggZZ},  except for the
 mixed chargino box contributions  arising
when  two different charginos are running along the box loop,
which  had not been computed before.
In Section 3 we compute the polarized
$\gamma\gamma\to ZZ$ differential cross section induced by the above
Higgs and background contributions. We discuss the shape of the
$ZZ$ invariant mass distribution and the observability of the
Higgs signal, in particular its dependence on the photon-photon flux
for polarized laser photon and  $e^{\pm}$ beams.
We show that, for a given Higgs mass, it is possible to optimize
the configuration by varying either the LC energy or the energy
of the laser beam. We give various  illustrations
with heavy SM and MSSM Higgs particles. The results are summarized
and commented in the concluding Section 4.

\section{The Higgs contribution to the $\gamma \gamma \to Z Z $
amplitudes.}

As  in \cite{ggZZ, gggg, gggZ}, we
use  the non-linear Feynman gauge   in which there are two
sets of diagrams contributing to
$\gamma \gamma \to ZZ$ \cite{Dicus}; compare Fig.\ref{diag-fig}.
The first consists of the one-particle irreducible "box"-diagrams
 involving  two photons and two $Z$'s as external legs;
see Fig.\ref{diag-fig}a.
Their contributions, arising from  loops involving  $W$'s
\cite{Jikia-ZZ}, quarks and leptons and
charginos \cite{Glover-ZZ}, as well as charged Higgs
particles and sfermions \cite{ggZZ}, are summarized
in the Appendix. We note in particular
that the "mixed" chargino boxes, induced by the
$Z\tchi_1 \tchi_2$ couplings involving  two different
charginos, are presented  in (\ref{F-chi1chi2-amp},
\ref{Amix1++++}-\ref{Amix2+-+-}). Numerically, they are not
expected to be particularly important. Nevertheless,
we list them here  for completeness,  because their derivation
(including their simplification to the present form)
required a considerable  effort. The analogous "mixed" sfermion
box contributions have not been calculated, since they should be
at most of  similar magnitude to the single sfermion,
which is already known to be very small \cite{ggZZ}.\par

The second set of diagrams\footnote{Notice that as opposed to the
  previous set, these are one-particle  reducible  diagrams.}
depicted in Fig.\ref{diag-fig}b,  consists of those
involving  contributions from a Higgs pole in the
$\s$-channel. These diagrams   contain
a Higgs-$\gamma \gamma$ vertex generated by
loops along which  $W$-gauge bosons (together with the associated
Goldstone and FP ghosts) and  fermion or physical scalar
particles are running. Their general form is
\bqa
&& F^h_{\lambda_1 \lambda_2 \lambda_3\lambda_4}(\gamma \gamma \to ZZ)
= -~\frac{\alpha^2}{2\swd\cwd}\Big \{  \H(\s)
\Big  \}
\cdot \nonumber \\ &&
\cdot \frac{(1+\lambda_1 \lambda_2)}{2}
~ \left [(1+\lambda_3 \lambda_4)\, \frac{\lambda_3 \lambda_4}{2}~ -
~\frac{1+\beta_Z^2}{1-\beta_Z^2}~ (1- \lambda_3^2)(1- \lambda_4^2)
\right ] \ , \label{Fh-pole}
\eqa\par
\noindent
where the hyperfine coupling is taken as  $\alpha=1/137$.

In SM, where only  one physical neutral Higgs particle exists with
mass  $m_H$, we have
\bq
\H(\s) =\sum_i N^c_iQ^2_i \F_i
(\tilde \tau)\frac{\s}{\s-m_H^2+i m_H \Gamma_H}  ~ , ~ \label{H-SM}
\eq
where  the index $i$ runs over the \underline{physical}
 charged particles  with spin (1,~ 1/2) running along the loop
with their  interactions determined by   \cite{Gunion-book}
\bq
\L_{H^0(SM)} =  g \mw W^+_\mu W^{\mu-} H^0 ~
-\frac{g m_f}{2 \mw} \bar \psi_f \psi_f H^0
- \frac{g m_{H}^2}{2\mw} G^+G^-H^0 \ , \label{h-vertex}
\eq
\noindent
where $G^{\pm}$ are the standard model Golstone bosons associated
to the $W^{\pm}$ bosons.
The W (plus Goldstone and FP ghost) and the charged
fermion  contributions to
$H^0\gamma \gamma$  are given respectively by
\bqa
\F_1(\tilde \tau) & = & \frac{2m_H^2}{\s} + 3 \tilde \tau +
3 \tilde \tau
\left ( \frac{8}{3} ~-~\frac{2 m_H^2}{3\s} ~
-\tilde \tau \right) f(\tilde \tau)
\ ,  \label{F1-SM} \\
\F_{1/2}(\tilde \tau) &=&
- 2\tilde \tau [1+ (1-\tilde \tau)f(\tilde \tau)] \ , \label{F12}
\eqa
where
\bq
\tilde \tau= ~\frac{4 m_i^2}{\s} ~~~~, ~~~~
f(\tilde \tau)=~-\frac{\s}{2} C_0(0,0, \s; m_i, m_i, m_i)
\ , \label{ftau}
\eq
with\footnote{A simple expression for $C_0$ in terms of logarithms
may be seen \eg in Eqs.(B.2) of \cite{ggZZ}.}
$C_0$ being the standard Passarino-Veltman function
\cite{Passarino} in the notation of \cite{ggZZ, Hagiwara};
(compare (\ref{BZ11}-\ref{DZZut})). In (\ref{H-SM}),
$Q_i$ is the charge  and $N^c_i$ the colour
multiplicity of the particle contributing to
$H^0\gamma \gamma $ loop.

The most important contributions to (\ref{H-SM}) in SM
come  from  $W^{\pm}$ (to which   Goldstone and FP ghost
are always included) and the  top-loop, which  are
determined  by (\ref{F1-SM}) and (\ref{F12}) respectively.\\ \par

It is also useful to remark that if a physical scalar
charged particle $H^\pm$ with mass $m_{H^\pm}$ were introduced
in SM interacting with the physical neutral Higgs as
\cite{Gunion-book}
\bq
\L_{H^0H^+H^-} =
- \frac{g m_{H^\pm}^2}{\mw} H^+H^-H^0 \ , \label{h-vertex-scalar}
\eq
then an additional contribution would arise in (\ref{H-SM})
determined by the function
\bq
\F_0(\tilde \tau)  = \tilde \tau
[1- \tilde \tau f(\tilde \tau)] \ , \label{F0}
\eq
which  is analogous to those in (\ref{F1-SM}, \ref{F12})
and  determines the contributions to
$H\gamma \gamma$  from spin=0 charged particles running along
the loop \cite{Gunion-book}.
Using this, it may then  be instructive to
notice  that the standard  $W^{\pm}$   contribution
 (\ref{F1-SM}) can   be written as
\bq
\F_1(\tilde \tau) = 3\tilde \tau
\left [1 +\left (\frac{8}{3}-\tilde \tau\right
)f(\tilde \tau) \right ]  +\frac{m_H^2}{2\mw^2}\F_0(\tilde \tau)
\ .  \label{FW-SM}
\eq
In the Feynamn gauge, the first term in (\ref{FW-SM}) gives the
pure $W$ and ghost SM contributions, while the last term (having
exactly the structure
that would be induced by a scalar charge=1 particle of mass $\mw$)
describes the Goldstone one. But of course,
such a separation is not   gauge invariant. \\ \par

In the MSSM case, with no
CP-violating phases in the new physics sector, there will be two
neutral CP-even Higgs particles $(h^0,~ H^0)$ contributing to
(\ref{Fh-pole}) so that
\bqa
\H(\s) &= &\sum_i N^c_iQ^2_i \Big [\sin(\beta-\alpha)\F^{h^0}_i
(\tilde \tau)\frac{\s}{\s- m_{h^0}^2+i m_{h^0} \Gamma_{h^0}}
\nonumber \\
&+ &\cos(\beta-\alpha)\F^{H^0}_i
(\tilde \tau)\frac{\s}{\s- m_{H^0}^2+i m_{H^0} \Gamma_{H^0}}
\Big ]  ~ , ~   \label{H-SUSY}
\eqa
where   the sum is  running over the
physical charged particles  with spin
(1, 1/2, 0) contributing to the $h^0\gamma \gamma$
and $H^0 \gamma \gamma $ vertices; these are $W^{\pm}$ (to which
$G^{\pm}$ and FP ghosts are always  included), as well as
charginos  $\tchi^{\pm}$, $H^{\pm}$ and sfermions $\tilde f_j$.
The  interaction Lagrangian determining the necessary couplings
is
\bqa
&& \L_{(h^0, H^0)(\rm SUSY)}=
g\mw [H^0 \cos(\beta-\alpha)+ h^0 \sin(\beta -\alpha)]
\Big \{ W^{-\mu}W^+_\mu +\frac{1}{2\cwd} Z^\mu Z_\mu
-H^+H^- \Big \}
\nonumber \\ &&
 + \frac{ g \mw}{2\cwd }  \cos 2\beta ~
[h^0 \sin(\alpha+\beta) - H^0 \cos(\alpha+\beta)  ]
( G^+G^{-} - H^+H^-)
\nonumber \\ &&
-\frac{g m_t}{2\mw \sbeta}[H^0\salpha +h^0 \calpha] \bar t t
-\frac{g }{2\mw \cbeta}[H^0\calpha- h^0 \salpha]
 ( m_b \bar b b +m_\tau \bar \tau \tau  )
\nonumber \\ &&
-\frac{g m_t^2}{\mw \sbeta}[h^0  \calpha
+ H^0  \salpha] (\Stop_1^*\Stop_1 +\Stop_2^* \Stop_2)
-\frac{g m_t}{2 \mw \sbeta}
 [h^0 (A_t \calpha +\mu \salpha)
\nonumber \\ &&
+ H^0 (A_t \salpha -\mu \calpha) ]
\sin(2\theta_t) \Sn(A_t-\mu \cot\beta)
(\Stop_1^*\Stop_1 -\Stop_2^* \Stop_2)
\nonumber \\ &&
-\frac{g\mw}{\cwd}  [H^0 \cos(\alpha+\beta)-
h^0 \sin(\alpha +\beta)]\Bigg \{
\Big [ \frac{2\swd}{3}+
 \Big (\frac{1}{2}- \frac{4\swd}{3} \Big )\cos^2 \theta_t
 \Big ]  \Stop_1^*\Stop_1
\nonumber \\ &&
+ \Big [ \frac{2\swd}{3}+
 \Big (\frac{1}{2}- \frac{4\swd}{3} \Big )\sin^2 \theta_t
 \Big ]  \Stop_2^*\Stop_2 \Bigg \}
\nonumber \\ &&
-\frac{g}{\sqrt{2}} \Del_1 \Big [
 \B_L \cfR \sfL (H^0 \calpha - h^0 \salpha)
+\B_R \sfR \cfL (H^0 \salpha + h^0 \calpha) \Big ]
\bar {\tchi}_1 \tchi_1
\nonumber \\ &&
 +\frac{g}{\sqrt{2}} \Del_2 \Big [
 \B_R \sfR \cfL (H^0 \calpha - h^0 \salpha)
\nonumber \\ &&
+\B_L \sfL \cfR (H^0 \salpha + h^0 \calpha) \Big ]
\bar {\tchi}_2 \tchi_2 ~ . \label{h-vertex-SUSY}
\eqa\par
The implied  $W$-loop  contributions  to the $h^0,~H^0$
terms in (\ref{H-SUSY}) are then respectively
\bqa
\F^{h^0}_1(\tilde \tau) & =& \sin(\beta-\alpha)~ 3\tilde \tau
\Big [1+\Big (\frac{8}{3}-\tilde \tau \Big ) f(\tilde \tau) \Big ]
-\frac{\cos(2\beta) \sin(\beta+\alpha)}{2\cwd} \F_0(\tilde \tau)
\ ,  \label{F1h-SUSY} \\
\F^{H^0}_1(\tilde \tau) & =& \cos(\beta-\alpha)~ 3\tilde  \tau
\Big [1+\Big (\frac{8}{3}-\tilde \tau \Big ) f(\tilde \tau) \Big ]
+\frac{\cos(2\beta) \cos(\beta+\alpha)}{2\cwd} \F_0(\tilde \tau)
\ ,  \label{F1H-SUSY}
\eqa
while the  contributions from  fermion and physical scalar particles
are given by  (\ref{F12}, \ref{F0}) respectively,
after substituting  of course  the obvious changes in the  couplings
implied by  comparing (\ref{h-vertex}, \ref{h-vertex-scalar}) with
 (\ref{h-vertex-SUSY}).

Concerning the parameters entering (\ref{h-vertex-SUSY}),
we   quote   the neutral Higgs  mixing angle
$\alpha$  determined by \cite{SUSY-reviews}
\bq
\tan (2\alpha )=\tan (2 \beta) \frac{m_{A^0}^2 +\mzd}
{m_{A^0}^2 -\mzd +\frac{\epsilon}{\cos(2 \beta)}}
~, \label{alpha-angle}
\eq
and the constraint   $-\pi/2 \leq \alpha \leq 0$.
The  leading top-stop contribution to
 (\ref{alpha-angle}) is    \cite{h0-mixing}
\bq
\epsilon \simeq \frac{3 G_F m_t^4}{\sqrt{2}\pi^2 \sin^2\beta}
 \ln \left (\frac{M_S^2}{m_t^2}\right )
 ~~ , \label{epsilon-mixing}
\eq
where $M_S^2 \simeq m_{\Stop_1}  m_{\Stop_2} $ provides a
measure of the SUSY breaking scale. \par

We also note that the neutral Higgs-$\Stop_i$
couplings in (\ref{h-vertex-SUSY})
depend on the $(\Stop_L,~ \Stop_R)$ mixing defined in
(\ref{stop-angle}, \ref{stop-angle-range}); while
 the neutral Higgs-chargino couplings
in (\ref{h-vertex-SUSY}), assume
the   mixing definition
in (\ref{chi-mass-matrix}-\ref{chi-angle-range})    and the
sign quantities  $\Del_1$, $\Del_2$, $\B_L$ and $\B_R$
given in   (\ref{signs}). The consistency of the formalism
guarantees
that the   chargino physical masses are always
positive, for any sign of $M_2$ and $\mu$.

\section{The $\gamma \gamma \to ZZ$ process close to the Higgs pole.}

We now explore the possibility of studying the contribution of a
Higgs $s$-channel pole using polarized
$\gamma\gamma$ collisions in an LC operated in the $\gamma
\gamma $ mode,  through laser backscattering
\cite{LCgg}. With Bose statistics and Parity invariance
 the general form for the
$\gamma \gamma \to ZZ$ cross section has been written in
\cite{ggZZ} for any polarization state of the photons, in terms of
helicity amplitudes. Here we
restrict to the case of circular laser polarization
(without any transverse linear part)   which turns
out to be
most interesting for the search of Higgs effects. The cross
section for the laser backscattered photons
normalized to unit electron positron flux is
then \cite{ggZZ}
\bqa
{d\sigma (\gamma \gamma \to ZZ)\over
d w ~d\cos\vartheta^*}\Bigg |_{\rm Laser}  &=&
\frac{1}{\sqrt{s_{ee}}}{d \bar L_{\gamma\gamma}\over d\rtau}
\Bigg \{ {d\bar{\sigma}_0\over d\cos\vartheta^*}
+\langle \xi_2 \xi_2^\prime \rangle
{d\bar{\sigma}_{22}\over d\cos\vartheta^*}
 \Bigg \} \ \ ,
\label{sigpol}
\eqa
where $w\equiv \Egg \equiv \sqrt{\s}$ is the c.m.
energy of the backscattered photons, equal to the $ZZ$ invariant mass, 
while $\sqrt{s_{ee}}$ is the $e^-e^+$ c.m. energy at which the LC is
operating. The  relevant $\gamma \gamma$ cross sections are
given by
\bqa
{d\bar \sigma_0(\gamma \gamma \to ZZ) \over d\cos\vartheta^*}&=&
\left ({\beta_Z\over 128 \pi\hat{s}}\right )
\sum_{\lambda_3\lambda_4} [|F_{++\lambda_3\lambda_4}|^2
+|F_{+-\lambda_3\lambda_4}|^2] ~ ,  \label{sig0} \\
{d\bar{\sigma}_{22}(\gamma \gamma \to ZZ)\over d\cos\vartheta^*} &=&
\left ({\beta_Z\over 128 \pi\hat{s}}\right )\sum_{\lambda_3\lambda_4}
[|F_{++\lambda_3\lambda_4}|^2
-|F_{+-\lambda_3\lambda_4}|^2]  \ , \label{sig22}
\eqa
in terms of helicity amplitudes defined in Appendix A.
In (\ref{sigpol}-\ref{sig22}), $\vartheta^*$ denotes
the Z-scattering angle in $\gamma \gamma$ c.m. frame.
 Note that  $d\bar \sigma_0/ d\cos\vartheta^*$ is the
unpolarized $\gamma \gamma \to ZZ $ cross section and therefore
it is positive definite,
while $d\bar \sigma_{22}/ d\cos\vartheta^*$ can be of either sign.
In terms of these, the cross section for case that both photons have
 helicity=+1, is expressed as
\bq
\frac{d\bar \sigma_{++}(\gamma \gamma \to ZZ)}{d\vartheta^*}~=~
\frac{d\bar \sigma_{0}(\gamma \gamma \to ZZ)}{d\vartheta^*}~+~
\frac{d\bar \sigma_{22}(\gamma \gamma \to ZZ)}{d\vartheta^*}~~ .
\label{sig++}
\eq \\ \par

  The quantity $d\bar L_{\gamma\gamma}/d\rtau$
in (\ref{sigpol}) describes the
photon-photon differential luminosity
per unit $e^-e^+$ (or $e^-e^-$)  flux at
$\tau \equiv s_{\gamma \gamma}/s_{ee}$; while
the Stokes parameters $\xi_2$, $\xi_2^\prime$ describe the
 helicities  of the two backscattered photons \cite{LCgg},
so that $\langle \xi_2 \xi_2^\prime \rangle $ denote the average
value of the product of these  helicities as a function of $\tau$.
Here, we follow  the same notation as in \cite{LCgg, Tsirigoti}
and the Appendix B of the last paper in \cite{gggg}. \par

As explained in \cite{LCgg},
after the Compton scattering of an electron beam of energy
$E$ from a laser photon of energy $\omega_0$,
the electron   looses most of its energy and a
beam of "backscattered photons" is
produced with the energy $\omega$, whose    fractional energy
 $x \equiv \omega/E$ satisfies
\bq
0 \leq x \leq x_{max} ~\equiv~ \frac{x_0}{1+x_0} \ \ \ \ ,
\ \ \ \ 0\leq x_0 \leq 2 (1+\sqrt 2) \ \ , \label{laser-kin1}
\eq
where $x_0=4 E \omega_0/m_e^2$. Applying laser backscattering
 to both electron
beams, we conclude that the c.m. energy of the produced
hard photons is constrained by
\bq
\tau ~ < ~ \frac{x_0 x_0^\prime}{(1+x_0) (1+x_0^\prime)} ~~,
\label{laser-kin2}
\eq
where we have allowed for the possibility that the energies of the
two laser photons may be different.\par

It turns out that the shapes of $d\bar L_{\gamma\gamma}/d\rtau$
and $\langle \xi_2 \xi_2^\prime \rangle $ strongly depend also
on the longitudinal polarizations of the two electron beams
$P_e$ and $P_e^\prime$ and on the average helicities of the
corresponding laser photons $P_\gamma$ and $P_\gamma^\prime$.
Examples of these are shown in Fig.\ref{spectra}a,b,
for the most interesting case $P_e=P'_e=0.8$,
$P_\gamma=P'_\gamma=-1$ in which both $d\bar L_{\gamma\gamma}/d\rtau$
and $\langle \xi_2 \xi_2^\prime \rangle$ peak at a $\tau$-value
close to the maximum allowed by (\ref{laser-kin2}).
Thus for the highest possible value of
$x_0=x_0^\prime=4.83$, this peak appears at $\rtau \simeq 0.8$
where $\langle \xi_2 \xi_2^\prime \rangle \simeq 1$.
 By varying $(x_0, ~x_0^\prime)$, for a given LC energy $E_e$,
it should be possible to move
the peak of these distributions at the value of the
mass of the Higgs boson one wants to study,
thereby increasing the sensitivity; compare Fig.\ref{spectra}a,b.
In Fig.\ref{spectra}c, the photon-photon Luminosity factor for
unpolarized electron and laser beams are given for comparison.
As discussed below, under certain circumstances it may be
advisable to use  such unpolarized beams in the Higgs searches!
\par

We now examine how the distribution in eq.(\ref{sigpol})
 can reflect the presence of a
Higgs boson and allow to study its properties.
In this respect the contributions from the diagrams of
Fig.\ref{diag-fig}b determine the "signal", while those of
Fig.\ref{diag-fig}a the  "$\gamma\gamma \to ZZ$-background".
\par

In a first step we fix the energy of the $e^+e^-$ LC to its maximal
value, for example at $500~GeV$ or  $800~GeV$ for the TESLA
project \cite{gg2000, Telnov-gg2000}.
For the SM Higgs boson case, we then look at the effect
for Higgs masses at  200, 300, 400 and 500 GeV\footnote{We note
  in passing that standard model Higgs masses of up to 500 GeV, or even
  larger, may easily be  made consistent with all
  experimental data and theoretical bounds, by either introducing new
  very heavy particles or extra large dimension. For a review see
  \cite{Quigg}.}
. As a second step,
depending on the situation, we can improve the sensitivity be
either optimizing the choice of the ($x_0$, $x_0^\prime$) values,
or the LC machine energy. Below, we discuss examples of both
situations.\\

\noindent
\underline{\bf SM cases:}.\\
As expected from the shapes of the
luminosity factors $d \bar L_{\gamma\gamma}/ d\rtau$
and $\langle \xi_2 \xi_2^\prime \rangle$ presented in
Fig.\ref{spectra}a,b, if $x_0=4.83$,
then the Higgs effect will be mostly visible in the case
of a Higgs mass corresponding to
$\sqrt{\tau_H} \equiv m_H/\sqrt{s_{ee}}\simeq 0.8$,
i.e. $m_H\simeq 400~\rm GeV$ for a $500~\rm GeV$ LC
machine.

We start therefore from this $m_H=400~\rm GeV$ case
presented in  Fig.\ref{SM-4-fig}, where the laser induced
cross section defined in (\ref{sigpol}) is plotted versus
$\rtau$. The total Higgs
width used in the calculation is
obtained from the code of  \cite{HDECAY} and it is indicated in
Fig.\ref{SM-4-fig}. In the same figure the
results for $m_H=100\rm GeV$ are also given as a help for estimating the
background. Note that the cross sections are always integrated in the
$30^0 <\vartheta^* <150^0$ angular region.\par

It is evident from Fig.\ref{SM-4-fig} that the
magnitude of the $m_H=400\rm GeV$ contribution, compared to the
background, is maximized in a region which is not symmetric
around the $\Egg=400~\rm GeV$ region. This must be related to
energy region where the signal-background  interference is most
constructive; as well as to the shape of the photon
spectrum and the magnitude of the Higgs width. \par

 Using (\ref{sigpol}), the expected number of events in
some appropriate  region of $ZZ$ invariant masses
$w_{\rm min} < w < w_{\rm max}$ around the Higgs pole
 is then defined  by
\bqa
 \N^{\rm SM} & =& \L_{ee}~ I_f
\int_{w_{\rm min}}^{w_{\rm max}} dw ~
\frac{d\sigma}{d w ~d\cos\vartheta^*}\Bigg |_{\rm Laser}
~~ , \label{SM-signal} \\
\N_{\rm Bg}^{\rm SM} & =& \L_{ee}~ I_f
\int_{w_{\rm min}}^{w_{\rm max}} dw ~
\frac{d\sigma (m_H=100 \rm GeV)}{ d w ~d\cos\vartheta^*}
\Bigg |_{\rm Laser}
~~ , \label{SM-background}
\eqa
for the signal and the background respectively.
Here $\L_{ee}$ is
the LC luminosity which for 0.5TeV TESLA LC is taken
$\L_{ee}=300 ~\rm fb^{-1}/year $ at its top
energy \cite{Telnov-gg2000}. From  (\ref{SM-signal},
\ref{SM-background}), the statistical significance of the effect
is then given  by
\bq
S.D. \equiv \frac{(\N^{\rm SM}- \N^{\rm SM}_{\rm Bg})}
{\sqrt{\N^{\rm SM}_{\rm Bg}}} ~ ~~. \label{SM-SD}
\eq\par

The quantity $I_f$ in (\ref{SM-signal}, \ref{SM-background}) denotes
the identification factor of the $Z$-pair. Assuming that the useful
modes for the ZZ identification are those where one Z decays
leptonically (including the invisible neutrino mode), and the other
hadronically, we get $I_f\simeq 0.47$; while if only charged leptons
are used  for the leptonic mode, $I_f\simeq 0.14$ is obtained.
In the following we will present event-numbers and statistical
significance corresponding to both  cases, $I_f=0.47$
and in parenthesis $I_f=0.14$.\par

Thus, for the SM case of $m_H=400 \rm GeV$, (presented in
Fig.\ref{SM-4-fig}) we get,  the
results indicated in the second column of Table 1.
Of course the results mildly depend also on the  choice of
$ZZ$ invariant masses $w_{\rm min}$
and $w_{\rm max}$ which, on the basis of Fig.\ref{SM-4-fig}, have
been taken asymmetrically around the Higgs mass. This choice
 is also indicated in Table 1. The conclusion thus reached is that
the  sensitivity of a 0.5 TeV LC to a $m_H=400\rm GeV$
 SM  Higgs  seems to be quite high.\par

\begin{table}[htb]
\begin{center}
{ Table 1: SM Higgs searches at a 0.5 TeV TESLA LC. \\
 ( $\L_{ee}=300 ~\rm fb^{-1}/year $ )}\\
\vspace*{0.3cm}
\begin{tabular}{||c|c|c|c|c|c||}
\hline \hline
\multicolumn{1}{||c|}{ } &
\multicolumn{1}{|c|}{ Fig. \ref{SM-4-fig} }&
\multicolumn{1}{|c|}{ Fig. \ref{SM-3-2-fig}a } &
\multicolumn{1}{|c|}{ Fig. \ref{SM-3-2-fig}b }&
\multicolumn{1}{|c|}{ Fig. \ref{SM-3-2-fig}c } &
\multicolumn{1}{|c||}{ Fig. \ref{SM-3-2-fig}d } \\
\hline
$m_H$ (GeV) & 400 & 300 &300 & 200 & 200  \\
$\Gamma_H$ (GeV)& 28.89 & 8.51 &8.51 & 1.428 & 1.428 \\
$x_0=x_0^\prime$ & 4.83 & 4.83 & 2 & 4.83 & 1 \\
$w_{\rm min} (GeV)$ & 340 & 290 & 290 & 195 & 195 \\
$w_{\rm max}$ (GeV)& 410 & 310 & 310 & 205 & 205 \\ \hline
$\N^{\rm SM}$ & 4635 & 807 & 3637 & 1033 & 4105 \\
 & (1381) & (240) & (1083) & (308) & (1223) \\ \hline
$\N^{\rm SM}_{\rm Bg} $ & 3188 & 250 & 574 & 568 & 705 \\
 &(950) &(75) & (171)  & (169) & (210) \\ \hline
S.D. &25.6 & 35 & 128 & 19.5 & 128 \\
 & (14) & (19) & (70) & (10.6) & (70) \\ \hline \hline
\end{tabular}
\end{center}
\end{table}

We next turn to lower Higgs masses.
If we insist using for them $x_0=x_0^\prime=4.83$ with the beams
polarized as in Figs.\ref{spectra}a,b
and $\sqrt{s_{ee}}=0.5 \rm TeV$,  then
the effect will be  weaker,
because of the smaller
values of the luminosity factor $d \bar L_{\gamma\gamma}/ d\rtau$
and of the polarization
 factor $\langle \xi_2 \xi_2^\prime \rangle$, which
may even become negative. As examples of such cases we give in
Figs.\ref{SM-3-2-fig}a,c the results for SM Higgs masses of
$m_H=300\rm GeV$ and $m_H=200\rm GeV$ respectively;
where \cite{HDECAY} has again been used for calculating
the needed Higgs total width.
The corresponding sensitivities are indicated in the columns of
Table 1 named  after  Fig.4a and Fig. 4c. As it is seen there,
the sensitivity is quite
considerable, in spite of the fact that
the background contribution due to the
($+-$) photon-photon helicity amplitude which does not contain
the Higgs effect, plays now a stronger role relatively
to the $(++$) one. We also note that the smallness of the
Higgs-width in these two cases, renders compelling   the symmetric
selection of $ZZ$ invariant masses
$w_{\rm min}$ and $w_{\rm max}$  around the value of
$m_H$, with $w_{\rm max}- w_{\rm min}<20 \rm GeV$; see Table 1. \par

The sensitivity to Higgs masses like those used in
Fig.\ref{SM-3-2-fig} can  be further increased by
reducing the energy $\omega_0$ of the laser, while still keeping
fixed the $e^+e^-$ energy. This way, the value of
$x_0$ may be reduced  so that the peak of the luminosity spectrum
shown in Fig.\ref{spectra}
coincides with $\sqrt{\tau_H} \equiv m_H/\sqrt{s_{ee}}$.
Thus, choosing  $x_0\simeq 1, ~2$ for LC(500), sets the photon  spectrum
peak at  $\sqrt{\tau_H} \simeq 0.4,~0.6$ respectively,
corresponding to $m_H\simeq 200,~300~\rm GeV$. The
results are indicated in Figs.\ref{SM-3-2-fig}b,d and the
corresponding columns  of Table 1. We see from them that the
sensitivity indeed improves a lot. Tuning therefore the
$x_0,~ x_0^\prime$-values for
such Higgs masses  may be a very rewarding idea.\par

Before ending the discussion  of SM masses  in the 200-300
GeV region, we should also  remark that if we insist in using
$x_0=x_0^\prime=4.83$ with  LC(500) running at its top energy,
then the employment  of polarized beams as
those indicated in Figs.\ref{spectra}a,b is  not particularly
appropriate. The reason is  that for such polarizations,
$\langle \xi_2 \xi_2^\prime \rangle$ is mostly
negative at the relevant $\sqrt{\tau_H}$ values.
Therefore, the results of  the columns labeled Fig.4a and Fig.4c in Table 1,
could  be improved by using instead unpolarized
electron and laser photons, for which
 $\langle \xi_2 \xi_2^\prime \rangle$
vanishes, and   the relevant
 $d \bar L_{\gamma\gamma}/ d\rtau$ values are  almost
$20\%$ larger; compare Fig.\ref{spectra}c.
Of course, these  improvements would not be as large  as those   induced
by the Fig.4b and Fig.4d choices in the same
table\footnote{These remarks are under the assumption
that the distance between the production and interaction points
of the backscattered photons is sufficiently
small; see the second paper in \cite{revggH}.}. \\ \par

We next turn to the case of  $m_H=500 \rm GeV$, as an example
of an SM Higgs mass which in the TESLA project
can only be studied in the 800~GeV Linear
Collider. For the machine luminosity in this case we use
  $\L_{ee}=500 ~\rm fb^{-1}/year $ at its top energy, while for
lower energies the luminosity is scaled down linearly
\cite{Telnov-gg2000}.\par

For the usual choice $x_0=x_0^\prime=4.83$ with the $e^\mp$ and laser
 polarizations indicated in Figs.\ref{spectra}a,b,
the results are
shown  in Fig.\ref{SM-5-fig}a, where the dash line gives the
prediction for $m_H=100\rm GeV$ and serves as an estimate of the
background. It is obvious from this figure, that in an LC(800) machine
 without some "tuning", it would not be possible to study such a high
mass SM Higgs. This remains true, even if we had used instead
unpolarized beams, as it can be inferred from the  flux
in Fig.\ref{spectra}c, and the results in Fig.\ref{SM-5-fig}a.\par

As a second attempt we assume $x_0=x_0^\prime=2$ for
 the polarized beam case, which  moves
the peaks of the  $d \bar L_{\gamma\gamma}/ d\rtau$ and
 $\langle \xi_2 \xi_2^\prime \rangle$ distributions at
$\Egg \simeq 500 \rm GeV$, for LC(800) running at its top energy
$\sqrt{s_{ee}}=0.8 \rm TeV$; compare Figs.\ref{spectra}a,b.
The corresponding results are  given in
Fig.\ref{SM-5-fig}b. Using then (\ref{SM-signal},
\ref{SM-background}, \ref{SM-SD}) with $w_{\rm min}=0.43 \rm TeV$
and $w_{\rm max}=0.52 \rm TeV$, we get $\N^{\rm SM}=5562~ (1657)$,
$\N^{\rm SM}_{\rm Bg}=5224~ (1556)$ for $I_f=0.47 ~(0.14)$.
The corresponding statistical significance of the effect
is then determined by ${\rm S.D.}=4.7 ~(2.5)$.\par

This effect can be  slightly increased by using instead
$x_0=x_0^\prime =4.83$ and tuning the energy
of the LC(800), so that $m_H/\sqrt{s_{ee}} \simeq 0.8$; \ie
at $\sqrt{s_{ee}}=0.63~\rm TeV$. The results
so obtained are shown  in Fig.\ref{SM-5-fig}c. Using the same
values for $w_{\rm min}$ and $w_{\rm max}$ and taking into account
the fact that the LC(800) luminosity scales linearly as
$\sqrt{s_{ee}}$ decreases \cite{Telnov-gg2000}, we get
$\N^{\rm SM}=5731~ (1707)$,
$\N^{\rm SM}_{\rm Bg}=5346~ (1592)$,
${\rm S.D.}=5.3 ~(2.9)$ for $I_f=0.47 ~(0.14)$.
It is worth remarking here that for such heavy and wide Higgs particles,
 tuning the LC energy is not much more efficient in
improving the signal, than tuning the laser energies.
\\ \par

We should also remark here that
the enhancements around the Higgs mass indicated
in Figs.\ref{SM-4-fig},\ref{SM-3-2-fig},\ref{SM-5-fig}
are  not only due
to the magnitude of the Higgs contribution sufficiently close
to its mass shell,
but also due to its  constructive interference with the predominantly
imaginary $F_{++++}$ amplitude induced by the $W$-loop
 \cite{ggZZ}. But as $m_H$ increases, this interference
decreases and eventually it  becomes destructive.
In fact, we have checked that exactly at $w= m_H$,
the interference  remains constructive
only for Higgs masses   $\lsim 460~ \rm GeV$.
Of course, off-shell there may be constructive
interference even for higher Higgs masses.
Because of this and the large SM Higgs width for sufficiently high
masses, it seems possible  to study through $\gamma \gamma \to ZZ$
standard Higgs masses up to $\simeq 500\rm GeV$;
compare Fig.\ref{SM-5-fig} and the above analysis .\par

Looking at the results presented above, we also  note the
fast fall-off of the sensitivity to SM Higgs particles, as the
Higgs mass increases.
This  can be understood from the $m_H$ dependence
of the quantity $Br(H\to\gamma\gamma)Br(H\to ZZ)/m^2_H$
 controlling the size of the Higgs contribution;
and from the rise of the $\sigma(\gamma\gamma\to ZZ)_{\rm Laser}$
background
 cross section with the energy; (compare  \cite{ggZZ}). For example,
$Br(H\to\gamma\gamma)Br(H\to ZZ)/m^2_H$ decreases by a factor of
$80$ as
$m_H$ increases from $300$ to $500~ GeV$ (mainly due to the
 decrease of $Br(H\to\gamma\gamma)$); while
the background cross section increases by a factor 2
for the corresponding energy rise; compare \eg
 \cite{Gunion-book, revggH}. These two effects
explain the strong decrease of the $S.D.$ number
as the Higgs mass increases.\\ \par

\noindent
\underline{\bf SUSY cases:}.\\
As an  application to SUSY, we  investigate
models in which   all soft SUSY breaking sfermion mass parameters
 are taken at the TeV scale and the stop  mixing
maximal  \cite{Djouadi-SUSY-group}. Such models have the tendency to
push  the mass of the CP even $h^0$ towards its highest possible
values \cite{HDECAY}. Thus, for a sufficiently heavy CP-odd
$A^0$ Higgs particle and a not very small $\tan\beta$,
they should be well within the presently
allowed region  \cite{Higgs-exp, Hagiwara-SUSY, Quigg}.
 Assuming also the gaugino unification condition
\[
~  M_1=\frac{5}{3} \tan^2\theta_W M_2 ~~,
\]
we  present  in the first three lines of Table 2
 five Sets of values
for the independent parameters $\mu$, $M_2$,  $M_{\tilde f}$,
 $\tan\beta$ and $A_t$. Taking also the mass $m_{A^0}$
of the CP-odd $A^0$ as a further  independent parameter,
we show  in  the same table  the implied stop, chargino and
Higgs masses, as well as the Higgs  widths and branching ratios
calculated using HDECAY \cite{HDECAY}.\par

We have  investigated $\gamma \gamma \to H^0 \to ZZ$ cases
where  $m_{A^0}$ is either 200 or
300 GeV, which imply similar (but somewhat higher masses) for the
CP-even $H^0$.
In all these cases, the $H^0$ particle we wish to study, has a width
of the order of 0.1 to 0.2 GeV. \par

Since this SUSY Higgs resonance is  much narrower than the typical
width of the peak of the photon-photon spectrum
(about 10~GeV for
${\rm LC(500)}_{\gamma \gamma}$); the expected number of events
within such a small region may  be written as
 \cite{LCgg,single-H0A0, Melles}
\bqa
\N^{\rm SUSY}  & \simeq & I_f ~ \L_{ee}
\frac{1}{\sqrt{s_{ee}}}\left (\frac{d \bar L_{\gamma\gamma}}{
d\rtau}\right )_{\tau=\tau_{H^0}} \cdot
\left \{ \Sigma^{\rm SUSY}_0
+\langle \xi_2 \xi_2^\prime \rangle _{\tau=\tau_{H^0}}
\cdot \Sigma^{\rm SUSY}_{22}  \right \} \ \ , \label{SUSY-signal}
\\
\N^{\rm SUSY}_{\rm Bg}  & \simeq &  I_f ~ \L_{ee}
\frac{1}{\sqrt{s_{ee}}} \left( \frac{d \bar L_{\gamma\gamma}}{
d\rtau}\right )_{\tau=\tau_{H^0}} \cdot
\left \{ \Sigma^{\rm Bg}_0
+\langle \xi_2 \xi_2^\prime \rangle _{\tau=\tau_{H^0}}
\cdot \Sigma^{\rm Bg}_{22}  \right \} \ \ , \label{SUSY-background}
\eqa
where
\bqa
\Sigma^{\rm SUSY}_0 &=& \int_{m_{H^0-5\rm GeV}}^{m_{H^0+5\rm GeV}}
d w ~ \bar \sigma_0(\gamma \gamma \to ZZ) ~~ , \label{Sigma0-SUSY}
\\
\Sigma^{\rm SUSY}_{22} &=& \int_{m_{H^0-5\rm GeV}}^{m_{H^0+5\rm GeV}}
d w ~ \bar \sigma_{22}(\gamma \gamma \to ZZ) ~~ , \label{Sigma22-SUSY}
\eqa
are expressed in terms of the
$\gamma \gamma \to ZZ$ subprocess  cross sections
defined in (\ref{sig0}, \ref{sig22}). As before,
$\tau_{H^0}=m_{H^0}^2/s_{ee}$. Notice that
(\ref{SUSY-signal}, \ref{SUSY-background}), can immediately be
derived from (\ref{SM-signal}, \ref{SM-background}) in the narrow
width approximation.
\par

As  examples of the form
of the $\bar \sigma_0(\gamma \gamma \to ZZ)$ and
$\bar \sigma_{22}(\gamma \gamma \to ZZ)$ cross sections in the SUSY case,
we show in Figs.\ref{SUSY-fig}a,b the results for the parameter
Sets    3 and 4 of Table 2. Quantities $\Sigma^{\rm SUSY}_0$
and  $\Sigma^{\rm SUSY}_{22}$ are then directly
calculated from them by integrating around the Higgs peak.
 The corresponding
background quantities $\Sigma^{\rm Bg}_0$ and  $\Sigma^{\rm Bg}_{22}$
are defined analogously to (\ref{Sigma0-SUSY}, \ref{Sigma22-SUSY})
by  subtracting  the resonance contributions to the
$\bar \sigma_0, ~\bar \sigma_{22}$ cross sections,
which typically have the structure
shown in Fig.\ref{SUSY-fig}. The values of  $\Sigma^{\rm SUSY}_0$,
$\Sigma^{\rm SUSY}_{22}$, $\Sigma^{\rm Bg}_0$,
$\Sigma^{\rm Bg}_{22}$ thus obtained,
for the cases of the five SUSY Sets
mentioned above, are  indicated at the end of Table
2.\par

\begin{table}[htb]
\begin{center}
{ Table 2: SUSY Sets. \\
(All running parameters are taken at the electroweak scale.) }\\
\vspace*{0.3cm}
\begin{tabular}{||c|c|c|c|c|c||}
\hline \hline
 & Set1 & Set2 & Set3 & Set4 & Set5 \\ \hline
\multicolumn{6}{||c||}{$M_2=200 \rm GeV$~ ,~ $\mu=300 \rm GeV$ ~,
 ~ $M_{\tilde f}=1000 \rm GeV $ }
\\ \hline
\multicolumn{1}{||c|}{$\tan \beta $ } &
\multicolumn{2}{|c|}{ 3 }&
\multicolumn{2}{|c|}{ 4 } &
\multicolumn{1}{|c||}{ 5 }
\\
\multicolumn{1}{||c|}{$A_t=A_b=A_\tau$ (GeV)}&
\multicolumn{2}{|c|}{2550 }&
\multicolumn{2}{|c|}{ 2600} &
\multicolumn{1}{|c||}{2550}
\\  \hline
\multicolumn{1}{||c|}{$m_{\Stop_1}$ (GeV)}&
\multicolumn{2}{|c|}{ 785 }&
\multicolumn{2}{|c|}{777 } &
\multicolumn{1}{|c||}{781 }
\\
\multicolumn{1}{||c|}{$m_{\Stop_2}$ (GeV)}&
\multicolumn{2}{|c|}{ 1198 }&
\multicolumn{2}{|c|}{1204}&
\multicolumn{1}{|c||}{1201}
\\
\multicolumn{1}{||c|}{$m_{\tchi_1} $(GeV) }&
\multicolumn{2}{|c|}{ 165 }&
\multicolumn{2}{|c|}{168 } &
\multicolumn{1}{|c||}{170}
\\
\multicolumn{1}{||c|}{$m_{\tchi_2}$ (GeV)}&
\multicolumn{2}{|c|}{ 340 }&
\multicolumn{2}{|c|}{339} &
\multicolumn{1}{|c||}{337}
\\ \hline
 $m_{A^0}$ (GeV) & 200  & 300  &200 &300 &200   \\
$m_{H^\pm}$ (GeV) & 214  & 310   & 215 & 310 &215   \\
$m_{h^0}$  & 109  & 113  &  115 & 118 &119  \\
$m_{H^0 }$ (GeV) & 212  & 307   & 207 & 304 &205  \\
$\Gamma_{H^0}$(GeV)  & 0.105  & 0.240   & 0.112 & 0.234 &0.135   \\
$Br(H^0 \to ZZ)$  & 0.188  & 0.069   & 0.123  &0.0475 &0.0747 \\
$Br(H^0 \to \gamma \gamma )\times 10^5 $
& 1.61   & 1.47 & 1.17  & 1.12 &0.769    \\ \hline \hline
$\Sigma_0^{\rm SUSY} $ (fb TeV)  & 0.99  & 0.64  & 0.64  & 0.58 & 0.39  \\
$\Sigma_{22}^{\rm SUSY} $ (fb TeV)  & 0.67  & 0.20   & 0.334  & 0.14 &0.09  \\ \hline
$\Sigma_0^{\rm Bg} $ (fb TeV) & 0.194  & 0.56  & 0.178  & 0.55 & 0.17  \\
$\Sigma_{22}^{\rm Bg} $ (fb TeV) & -0.122  & 0.11   & -0.125  & 0.11 &-0.124 \\ \hline
 & Set1 & Set2 & Set3 & Set4 & Set5   \\ \hline \hline
\end{tabular}
\end{center}
\end{table}

In analogy to (\ref{SM-SD}), the $H^0$-sensitivities in the
SUSY case are determined from
(\ref{SUSY-signal}, \ref{SUSY-background}) and the
results in Table 2, through
\bq
S.D. \equiv \frac{(\N^{\rm SUSY}- \N^{\rm SUSY}_{\rm Bg})}
{\sqrt{\N^{\rm SYSY}_{\rm Bg}}} ~ ~~. \label{SUSY-SD}
\eq
These sensitivities depend  of course also on the LC luminosity and
the parameters
\[
\left (\frac{d \bar L_{\gamma\gamma}}{
d\rtau}\right ) _{\tau=\tau_{H^0}} ~~~,
~~~ \langle \xi_2 \xi_2^\prime \rangle _{\tau=\tau_{H^0}} ~~ ,
\]
 which in turn are determined by the LC-energy and the
$x_0,~ x_0^\prime$ values and polarizations used.
For definiteness we assume a
TESLA LC(500), with a $\L_{ee}=300 \rm fb^{-1}/year$, at the top of
its energy \cite{Telnov-gg2000}. We then discuss below
the sensitivities   for each of the five Sets of
parameters in Table 2, by considering in each case,
four  different choices of the machine parameters.
In the first three choices we use the electron and laser polarizations
appearing in Figs.\ref{spectra}a,b; while in  the fourth choice,
the unpolarized beams inducing the solid line prediction in
Fig.\ref{spectra}c are used.
We first list
the results for the four  choices concerning Set 1,  in which
$m_{H^0}\simeq 212 ~\rm GeV$ and $\tan\beta=3$. They are
\begin{itemize}
\item
\underline{Set 1, Choice 1}.
LC(500) is run at $\sqrt{s_{ee}}=0.5 \rm TeV$ using
$x_0=x_0^\prime=4.83$. The electrons (positrons) and the laser
photons are taken polarized with polarizations
$P_e=P_e^\prime= 0.8$, $P_\gamma=P_\gamma ^\prime= -1$.
For the $m_{H^0}$-value of Set 1,
we then find
$\sqrt{\tau_{H^0}}=0.42$, implying from Fig.\ref{spectra}a,b
\[
\left(\frac{d \bar L_{\gamma\gamma}}{
d\rtau}\right )_{\tau=\tau_{H^0}}=1.25 ~~,
~~~ \langle \xi_2 \xi_2^\prime \rangle _{\tau=\tau_{H^0}}=-0.52 ~,
\]
which through (\ref{SUSY-signal}, \ref{SUSY-background}
\ref{SUSY-SD}) gives
\[
\N^{\rm SUSY}=226 ~(67.4)~~, ~~
\N^{\rm SUSY}_{\rm Bg}=91 ~(27)~~, ~~ S.D.=14.2~(7.8) ~~
\]
 for $I_f= 0.47~(0.14)$ respectively.
\item
\underline{Set 1, Choice 2}.
LC(500) still  runs at $\sqrt{s_{ee}}=0.5 \rm TeV$,  but
$x_0=x_0^\prime=1$ is now used, which for the
$m_{H^0}$-value of Set 1 again gives
$\sqrt{\tau_{H^0}}=0.42$, implying from Fig.\ref{spectra}a,b
\[
\left (\frac{d \bar L_{\gamma\gamma}}{
d\rtau}\right )_{\tau=\tau_{H^0}}=1.89 ~~,
~~~ \langle \xi_2 \xi_2^\prime \rangle _{\tau=\tau_{H^0}}=0.70 ~.
\]
Through (\ref{SUSY-signal}, \ref{SUSY-background}
\ref{SUSY-SD}), this gives
\[
\N^{\rm SUSY}=778 ~(232)~~, ~~
\N^{\rm SUSY}_{\rm Bg}=57.9 ~(17.2)~~, ~~ S.D.=94~(51) ~~
\]
 for $I_f= 0.47~(0.14)$ respectively.
\item
\underline{Set1, Choice 3}.
This is the "extreme tuning" case in which $x_0=x_0^\prime=4.83$
is used, and LC(500) is tuned  at $\sqrt{s_{ee}}=0.265~ \rm TeV$, so that
for the $m_{H^0}$-value of Set 1 we are guaranteed to have
$\sqrt{\tau_{H^0}}\simeq 0.8$, implying from Fig.\ref{spectra}a,b
\[
\left (\frac{d \bar L_{\gamma\gamma}}{
d\rtau}\right )_{\tau=\tau_{H^0}}=2 ~~,
~~~ \langle \xi_2 \xi_2^\prime \rangle _{\tau=\tau_{H^0}}=0.92 ~.
\]
Through (\ref{SUSY-signal}, \ref{SUSY-background}
\ref{SUSY-SD}) we then get
\[
\N^{\rm SUSY}=906 ~(270)~~, ~~
\N^{\rm SUSY}_{\rm Bg}=46.1 ~(13.7)~~, ~~ S.D.=127~(69) ~~
\]
 for $I_f= 0.47~(0.14)$ respectively.
\item
\underline{Set 1, Choice 4}.
LC(500) now   runs again at $\sqrt{s_{ee}}=0.5 \rm TeV$,  with
$x_0=x_0^\prime=4.83$, but unpolarized electron and laser beams
are used. Thus   for  the
$m_{H^0}$-value of Set 1 implying
$\sqrt{\tau_{H^0}}=0.42$, we get from Fig.\ref{spectra}c
\[
\left (\frac{d \bar L_{\gamma\gamma}}{
d\rtau}\right )_{\tau=\tau_{H^0}}=1.49 ~~,
~~~ \langle \xi_2 \xi_2^\prime \rangle _{\tau=\tau_{H^0}}=0 ~.
\]
Through (\ref{SUSY-signal}, \ref{SUSY-background}
\ref{SUSY-SD}), this gives
\[
\N^{\rm SUSY}=415 ~(124)~~, ~~
\N^{\rm SUSY}_{\rm Bg}=81.5 ~(24.3)~~, ~~ S.D.=79~(20) ~~
\]
 for $I_f= 0.47~(0.14)$ respectively.
\end{itemize}

Making a similar treatment for the cases of the other SUSY Sets of
Table 2, we obtain for Sets 3 and 5 concerning an $H^0$  in the
mass region of 200 GeV, the results in Table 3; while in Table 4
we give the results for Sets 2 and 4 concerning $m_{H^0}\sim 300
~\rm GeV$. As expected from the SM discussion
 about the 200-300 GeV mass region, the
sensitivities in the case of  Choice 4 (employing  unpolarized beams)
are always better than those of Choice 1, but worse than those of
Choices 2 and 3.

\begin{table}[htb]
\begin{center}
{ Table 3: SUSY Sets 3 and 5 at LC(500),\\
(Cases with $m_{H^0}\sim 200 ~\rm GeV $ for $I_f=0.47~(0.14)$.)  }\\
\vspace*{0.3cm}
\begin{tabular}{||c|c|c|c|c|c|c|c|c||}
\hline \hline
\multicolumn{1}{||c|}{ } &
\multicolumn{4}{|c|}{Set 3  } &
\multicolumn{4}{|c||}{Set 5  }
\\ \hline
\multicolumn{1}{||c|}{$\tan \beta$ } &
\multicolumn{4}{|c|}{4  } &
\multicolumn{4}{|c||}{5  }
\\ \hline
\multicolumn{1}{||c|}{Choice } &
\multicolumn{1}{|c|}{ 1 } &
\multicolumn{1}{|c|}{ 2 } &
\multicolumn{1}{|c|}{ 3 } &
\multicolumn{1}{|c|}{ 4 } &
\multicolumn{1}{|c|}{ 1 } &
\multicolumn{1}{|c|}{ 2 } &
\multicolumn{1}{|c|}{3 } &
\multicolumn{1}{|c||}{ 4 }  \\ \hline
$\sqrt{s_{ee}}$ (TeV) &0.5 &0.5 & 0.259& 0.5 & 0.5 & 0.5& 0.256& 0.5\\
$x_0=x_0^\prime$ &4.83 & 1 &4.83 & 4.83 & 4.83& 1 & 4.83 & 4.83\\
$\sqrt{\tau_{H^0}}$ & 0.414 &0.414 &0.8 &0.414 & 0.41 &0.41 &0.8 & 0.41\\[0.2cm]
$\left (\frac{d \bar L_{\gamma\gamma}}{d\rtau}\right )_{\tau_{H^0}}$
& 1.25 & 1.89 & 2 &1.49 & 1.26 &1.86 & 2 & 1.49\\[0.2cm]
$ \langle \xi_2 \xi_2^\prime \rangle _{\tau_{H^0}} $
& -0.51& 0.68 & 0.92 & 0 &-0.5 & 0.6& 0.92 & 0 \\[.3cm]
$\N^{\rm SUSY}$ & 165 & 462 & 534 & 269 & 122 & 233 & 267 & 164\\
&(49) &(138) &(159) & (80.1) &(36.2) &(69.4) & (79.4) & (48.8) \\[0.3cm]
$\N^{\rm SUSY}_{\rm Bg} $ & 85 &66 &36 & 75 & 81.4 & 49.6 & 31 & 71 \\
&(25) &(15) & (11) & (22.3) &(24.3) &(14.8) & (9.2) & (21.1) \\ [0.3cm]
S.D. & 8.7 &49 & 84 &22 & 4.5 &26 & 42 &11 \\
&(4.8) & (32) &(45) & (12) &(2.4) & (14) & (23) & (6) \\ \hline \hline
\end{tabular}
\end{center}
\end{table}
\begin{table}[htb]
\begin{center}
{ Table 4: SUSY Sets 2 and 4 at LC(500),\\
(Cases with $m_{H^0}\sim 300 ~\rm GeV $ for $I_f=0.47~(0.14)$ ).  }\\
\vspace*{0.3cm}
\begin{tabular}{||c|c|c|c|c|c|c|c|c||}
\hline \hline
\multicolumn{1}{||c|}{ } &
\multicolumn{4}{|c|}{Set 2  } &
\multicolumn{4}{|c||}{Set 4  }
\\ \hline
\multicolumn{1}{||c|}{$\tan \beta$ } &
\multicolumn{4}{|c|}{ 3  } &
\multicolumn{4}{|c||}{ 4  }
\\ \hline
\multicolumn{1}{||c|}{Choice } &
\multicolumn{1}{|c|}{ 1 } &
\multicolumn{1}{|c|}{ 2 } &
\multicolumn{1}{|c|}{ 3 } &
\multicolumn{1}{|c|}{4 } &
\multicolumn{1}{|c|}{ 1 } &
\multicolumn{1}{|c|}{ 2 } &
\multicolumn{1}{|c|}{3 } &
\multicolumn{1}{|c||}{4 } \\ \hline
$\sqrt{s_{ee}}$ (TeV) &0.5 &0.5 & 0.384 &0.5 & 0.5 & 0.5& 0.38 &0.5 \\
$x_0=x_0^\prime$ &4.83 & 2 &4.83& 4.83 & 4.83& 2 & 4.83 &4.83 \\
$\sqrt{\tau_{H^0}}$ & 0.613 &0.613 &0.8 &0.613 & 0.608 &0.608 &0.8&0.613 \\[0.2cm]
$\left (\frac{d \bar L_{\gamma\gamma}}{d\rtau}\right )_{\tau_{H^0}}$
& 1.09 & 1.73 & 2 & 1.29 & 1.09 &1.74 & 2 & 1.29\\[0.2cm]
$ \langle \xi_2 \xi_2^\prime \rangle _{\tau_{H^0}} $
& -0.4 & 0.88 & 0.92 &0 &-0.4 & 0.85 & 0.92 & 0 \\[.3cm]
$\N^{\rm SUSY}$ & 173 & 398 & 465 & 231 & 162 & 341 & 398 & 211 \\
&(51.6) &(119) &(138)&(68.6) &(48.2) &(102) & (118) & (62.7)\\[0.3cm]
$\N^{\rm SUSY}_{\rm Bg} $ & 159 &320 &373 & 202& 156 & 315 & 367 & 200 \\
&(47.5) &(95.4) & (111)&(60.2) &(46.6) &(94) & (109) & (59.7) \\ [0.3cm]
S.D. & 1.1 &4.3 & 4.7 &2.0 & 0.41 &1.4 & 1.6 & 0.7\\
&(0.6) & (2.4) &(2.6)& (1.1) &(0.23) & (0.79) & (0.87) & (0.4)\\ \hline \hline
\end{tabular}
\end{center}
\end{table}

As we see from  Tables 2-4, the number of standard deviations S.D. of the
signal is largest in the lower side of $\tan \beta$ and $A^0$ (or
$H^0$ ) masses considered
(\ie for $\tan\beta \simeq 3$ and $m_{A^0} \simeq 200 $GeV).
But as either $\tan\beta$ or $m_{A^0}$ increase, S.D. is diminishing
rather quickly. For $m_{A^0}=200$ GeV, it is still sizable even for
$\tan\beta =5$, provided we tune the  $x_0,~ x_0^\prime$-values at least.
\par

 For higher $H^0$ masses, the situation
becomes more difficult. Thus the sensitivity to
 $m_{H^0}\sim 300 ~\rm GeV $ of an LC(500) TESLA machine can only reach
some  modest levels, provided that
$\tan \beta \lsim 3$ and that  $x_0,~ x_0^\prime$ or energy tuning
are applied; Compare the results
in Tables 2 and 4. \par

We also note that while for $m_{H^0}\sim 200~\rm GeV$, LC-energy
tuning seems to be  more efficient than  $x_0,~ x_0^\prime$ tuning
for improving the signal; they become comparable for
$m_{H^0}\sim 300~\rm GeV$.

In order to understand the dependence in the Higgs mass of the
observability of the above SUSY cases,
one should not only consider, as in the SM case,
the quantity  $Br(H\to\gamma\gamma)Br(H\to ZZ)/m^2_H$
and  the energy rise of the
background cross sections  $\sigma_{0}(\gamma \gamma \to ZZ)$ and
$\sigma_{22}(\gamma \gamma \to ZZ)$; but also
the fact that the $H^0$ is narrow, its width being much smaller than
the width of the peak of the photon-photon spectrum
 ($\Delta \simeq 10~\rm GeV$).
This last effect leads to a reduction
of the signal by a factor $(\Gamma_{H^0}/\Delta)$; (in the SM case
this effect does not occur for  $m_H\gsim300~ \rm GeV$, because
$\Gamma_H \gsim10~ \rm GeV$). Using then the $\Gamma_H$ results
given in Table 2,  one can  easily
understand the values of the corresponding
S.D. numbers.\par

We next briefly comment on the interference pattern between the
$H^0$-pole and the  background contribution in the SUSY case.
Such an effect would be  evident  in Fig.\ref{SUSY-fig}b,
which corresponds to $m_{H^0}=300$ GeV and $\Gamma_{H^0}=0.234$ GeV,
provided that the energy resolution were perfect. But no
interference pattern is obvious in the  $m_{H^0}=200$ GeV case
of Fig.\ref{SUSY-fig}a.
Of course, with an energy resolution of about 10 GeV
(as we have assumed in our analysis, due to the peak of the
photon-photon spectrum) it is not possible to observe
  interference patterns of the type of Fig.\ref{SUSY-fig}b,
by just averaging the data symmetrically
around the Higgs mass. Nevertheless, it is
important to remember  that they might exist
and to search for them by trying various  ways  of  bining the
experimental data. In fact it is true  that
selections might exist which  could  appreciably modify
the number of signal events, (say \eg by selecting events
mainly  on one side
of the resonance we are searching for) and thus
help revealing a Higgs effect.\par

\section{Conclusions}

Assuming that the $\gamma \gamma $ Collider will be realized some day
by applying the laser backscattering idea  to  an $e^+e^-$ or $e^-e^-$
LC, we have studied the observability of a standard or SUSY  heavy
neutral Higgs boson produced in the $s$-channel through
$\gamma\gamma \to H^0 \to ZZ$. One of the motivations for performing
this work was to investigate  whether we could exploit the
striking  predominance of the
helicity conserving purely imaginary amplitudes expected
for the background
$\gamma \gamma \to ZZ$ process at sufficiently high energies.\par

We have considered both, the case of the SM  Higgs boson,
as well as  cases  for  the  heavier
CP-even $H^0$ Higgs  predicted in  MSSM. Under such circumstances,
we have computed the  amplitude for $\gamma\gamma\to H^0 \to ZZ$
in which the $\gamma\gamma\to H^0 $ coupling arises at one-loop, while
$H^0\to ZZ$ exists at tree level;
as well as the background $\gamma\gamma\to ZZ$ contribution
arising through one-loop box amplitudes.  In Section 2 and
the Appendix, we have collected
the explicit  analytic expressions for the Higgs-pole and background
amplitudes  in the SM and  the MSSM cases. They are presented
in terms of Passarino-Veltman functions
immediately suitable for computation. For most of the
formulae we refer to  \cite{Glover-ZZ, Jikia-ZZ, ggZZ},
except for  the mixed chargino box contributions
arising from two different charginos running along the box diagram,
which appear  here for the first time.\par

We have then studied the interplay of the resonant Higgs contribution
with the predominantly helicity conserving   imaginary
background amplitude at sufficiently high  energy.
Depending on the parameters of the model, remarkable
interference effects may appear, which in
some  cases enhance the Higgs signal. \\

Our  first application has been  to  the standard Higgs search.
For LC(500), using the photon-photon spectrum implied by the
highest meaningful laser energy ($x_0=4.83$) with circular photon
polarizations and longitudinally polarized $e^{\pm}$ beams, we
have shown what would be  the signal of an SM Higgs boson of
arbitrary mass, above the $ZZ$-threshold. In these illustrations we
have assumed the machine to be running at its top energy  of $0.5
\rm ~TeV$.

Provided the  Higgs mass is known, it should also be possible to
optimize the signal;  either by changing the energy
of the laser for fixed
$e^+e^-$ energy, or by tuning the LC energy. We have made
illustrations for $m_H=200,~300,~400~\rm GeV$ at LC(500) and for
$m_H=500~\rm GeV$ at LC(800), using the TESLA luminosities.\par

We find
that for   masses in the region ($m_H \simeq 200~-~300~\rm GeV$),
the narrow Higgs peak largely dominates the background, and the
interference effect does not play an important role; compare
Fig.\ref{SM-3-2-fig}.  For such Higgs masses, we have also observed that
when running  LC(500) at its top energy, the relevant values of
$\langle \xi_2 \xi_2^\prime \rangle$ are largely negative,
thus reducing the signal. Thus, under such energy running conditions
and Higgs masses, the use of  unpolarized electron and laser beams
will be more efficient than the use of the polarized beams mentioned
above. Of course, when $x_0, ~x_0^\prime$ or energy tuning is
employed, the importance of polarizations is re-established. \par

 For higher masses ($400~-~450~\rm GeV$),
the constructive interference between the large imaginary
parts of the Higgs and of the background amplitudes, increases the
size of the Higgs contribution. This is stronger in the region
just below the Higgs peak (compare Fig.\ref{SM-4-fig}),  and it
appears even for higher masses (see  Fig.\ref{SM-5-fig}b,c).
In such cases, the interference pattern plays an important role
for the Higgs detection.\par

Therefore,   the measurement of the various terms of the polarized
cross section $\bar \sigma_j (\gamma  \gamma \to Z Z)$,
constitutes a useful tool of  the standard Higgs
search, for Higgs masses in the range ($ 2\mz \lsim m_H \lsim
500~\rm GeV$). Compare the results in Table 1; and those for
$m_H\simeq 500~\rm GeV$ at LC(800), provided that either
laser energy or the LC energy is appropriately tuned.

For even  higher Higgs masses though, the strongly
decreasing $Br(H\to\gamma\gamma)$ branching ratio prohibits  an
observable effect. \\ \par

We have then turned to SUSY and  considered the effects of the heavier
CP-even $H^0$  predicted in MSSM.
The search has been concentrated on the  part of  the
 MSSM parameter space in which all sfermion SUSY breaking masses
and their mixing are  sufficiently large, in order to guarantee that
$m_{h^0}$ is  well within the presently allowed region.
We have considered five such Sets of SUSY parameters leading to
$m_{H^0}$ in either the 200GeV or the 300GeV mass region;
compare Table 2.  In all such cases, the SUSY
$H^0$ boson differs from the  SM one,
by having a much narrower width
and a smaller  branching ratio to  $ZZ$.

For the observability of the $H^0$ predicted in each of the
above parameter Sets,  we have studied, four choices of LC running
conditions; compare Tables 3 and 4.
In the first three choices, polarized $e^\mp$ and laser beams are used with
polarizations $P_e=P_e^\prime=0.8$ and $P_\gamma
=P_\gamma^\prime=-1$ respectively; while in the fourth choice
fully unpolarized beams were employed. As in the SM case, we have found,
that when running LC(500) at its top energy with $x_0=4.83$, the use of
unpolarized beams for studying Higgs masses in the 200-300GeV region
is more advantageous, than the use of the aforementioned
polarizations. But, the importance of
polarization is re-established when either $x_0, ~ x_0^\prime$ or
energy tuning is employed in order to improve the signal.

Our overall conclusion is that  the
observability limit in SUSY decreases to about $m_{H^0} \sim 300~GeV$
for $\tan \beta \lsim 4$; while for $\tan \beta \simeq 5$ it
goes down to almost 200 GeV; see Tables 2-4.
 In the computations  we have of course
 taken into account the SUSY box
contributions to $\gamma\gamma\to ZZ$ background.\par

We should also remark on the basis of the
various numerical investigations performed,  that
the interference pattern between the Higgs-pole and background
contributions varies, depending on the values of the SUSY
parameters used. Examples of such an effect may be seen
in  Figs.\ref{SUSY-fig}a,b.
This is similar to what has also been observed
in the SM case above. For comparison, in all
SM cases we have also given the
results for $m_H=100~\rm GeV$. \\ \par

In conclusion we can say that an experimental investigation
of the $\gamma\gamma\to ZZ$ process and an  analysis taking into
account the interference between  the Higgs resonance and the one loop
background amplitudes, should be helpful for the
identification of a scalar Higgs-like candidate.  \\

{\bf Acknowledgments.}\\
 We are pleased to thank Abdelhak  Djouadi for very
useful and informative  discussions about the SUSY Higgs properties.
One of us (GJG) would also like to thank the CERN Theory Division for
the   hospitality extended to him while
part of this work was performed.

\vspace{3cm}

\newpage

\renewcommand{\theequation}{A.\arabic{equation}}
\renewcommand{\thesection}{A.\arabic{section}}
\setcounter{equation}{0}
\setcounter{section}{0}

{\large \bf Appendix: The $\gamma \gamma \to Z Z $
amplitudes in the Standard and SUSY models.}

The invariant helicity amplitudes for  the process
\bq
\gamma (p_1,\lambda_1) \gamma (p_2,\lambda_2) \to
Z (p_3,\lambda_3) Z (p_4,\lambda_4) \ \ ,
\label{ggZZ-process}
\eq
are denoted as\footnote{The same definitions as in
\cite{ggZZ} are used.}
$F_{\lambda_1 \lambda_2 \lambda_3\lambda_4}(\beta_Z,\t,\u)$,
where the momenta and
helicities of the incoming  photons and outgoing $Z$'s
 are indicated in parentheses, and
\bq
\s=(p_1+p_2)^2 = \frac{4\mzd}{1-\beta_Z^2}
~~ ~,~ ~ \t=(p_1-p_3)^2 ~ ~,~ ~\u=(p_1-p_4)^2 ~ ,
\label{kin1}
\eq
\bq
\s_4=\s-4\mzd ~, ~ \s_2=\s-2\mzd ~, ~ \t_1=\t-\mzd
~,~ \u_1=\u- \mzd ~  ~ \label{kin2}
\eq
are used. Denoting by $\vartheta^*$
the c.m. scattering angle of $\gamma \gamma \to ZZ$,
we also note
\bqa
\t=\mzd -\frac{\s}{2}(1-\beta_Z \cos\vartheta^*) & , &
\u=\mzd -\frac{\s}{2}(1+\beta_Z \cos\vartheta^*) ~~ ,
\label{kin3} \\
Y=\t \u -\mz^4=\frac{s^2\beta_Z^2}{4}  \sin^2 \vartheta^*=\s p_{TZ}^2
& , & \Delta =\sqrt{\frac{\s \mzd}{2Y}} ~ ,\label{kin4}
\eqa
where $p_{TZ}$ is the $Z$ transverse momentum.

The parameter   $\beta_Z$ in (\ref{kin1})
coincides with
the $Z$-velocity in the $ZZ$ c.m. frame. As in
\cite{Jikia-ZZ, Glover-ZZ, ggZZ} it is used
instead of  $\s$. The
standard form of the $Z$ polarization vectors implies
the useful relation
\bq
F_{\lambda_1 \lambda_2 \lambda_3\lambda_4}(\beta_Z,\t,\u) =
F_{\lambda_1 \lambda_2, -\lambda_3, -\lambda_4}(-\beta_Z,\t,\u)
(-1)^{\lambda_3-\lambda_4} \ , \label{-beta}
\eq
among the various helicity amplitudes. In addition,
Bose statistics, combined with the Jacob-Wick (JW)
\cite{JW-convention}
phase conventions   demands
\bqa
F_{\lambda_1 \lambda_2 \lambda_3\lambda_4}(\beta_Z,\t,\u) &=&
F_{\lambda_2 \lambda_1 \lambda_4\lambda_3}(\beta_Z,\t,\u)
(-1)^{\lambda_3-\lambda_4} \ , \label{Bose2} \\
F_{\lambda_1 \lambda_2 \lambda_3\lambda_4}(\beta_Z,\t,\u) &=&
F_{\lambda_2 \lambda_1 \lambda_3\lambda_4}(\beta_Z,\u,\t)
(-1)^{\lambda_3-\lambda_4} \ , \label{Bose1} \\
F_{\lambda_1 \lambda_2 \lambda_3\lambda_4}(\beta_Z,\t,\u) &=&
F_{\lambda_1 \lambda_2 \lambda_4\lambda_3}(\beta_Z,\u,\t)
 \ . \label{Bose12}
\eqa
while  CP invariance, being equivalent to parity invariance
at the 1-loop level, implies\footnote{A sign error in Eqs.(A.13)
of \cite{ggZZ} is corrected in (\ref{-beta++-0}) here.}
\bq
F_{\lambda_1 \lambda_2 \lambda_3\lambda_4}(\beta_Z ,\t,\u) =
F_{-\lambda_1,-\lambda_2,- \lambda_3,-\lambda_4}(\beta_Z,\t,\u)
(-1)^{\lambda_3-\lambda_4} \  .  \label{parity}
\eq

Using (\ref{-beta}) we remark that
\bqa
F_{++--}(\beta_Z, \t,\u) & = & F_{++++}(-\beta_Z, \t, \u)
\  \ , \label{-beta++--} \\
F_{++-0}(\beta_Z, \t,\u) & = & -F_{+++0}(-\beta_Z, \t, \u)
\  \ ,  \label{-beta++-0}
\eqa
which combined with $(\t \leftrightarrow \u)$
or helicity changes  and the use of
 (\ref{Bose2}-\ref{parity}), allow to express  the
36 $\gamma \gamma \to ZZ $ helicity amplitudes
in terms of just the eight independent ones
\cite{ggZZ, Jikia-ZZ, Glover-ZZ}
\bqa
& F_{+++-}(\beta_Z,\t,\u)~ , ~ F_{++++}(\beta_Z,\t,\u), &
  F_{+-++}(\beta_Z,\t,\u)~, ~ F_{+-00}(\beta_Z,\t,\u)~, \nonumber \\
&F_{++00}(\beta_Z,\t,\u)~ ,~  F_{+++0}(\beta_Z, \t,\u), &
F_{+-+0}(\beta_Z,\t,\u)~ , ~  F_{+-+-}(\beta_Z,\t,\u )~.
\label{8basic}
\eqa  \par

\begin{figure}[htb]
\vspace*{3cm}
\[
\epsfig{file=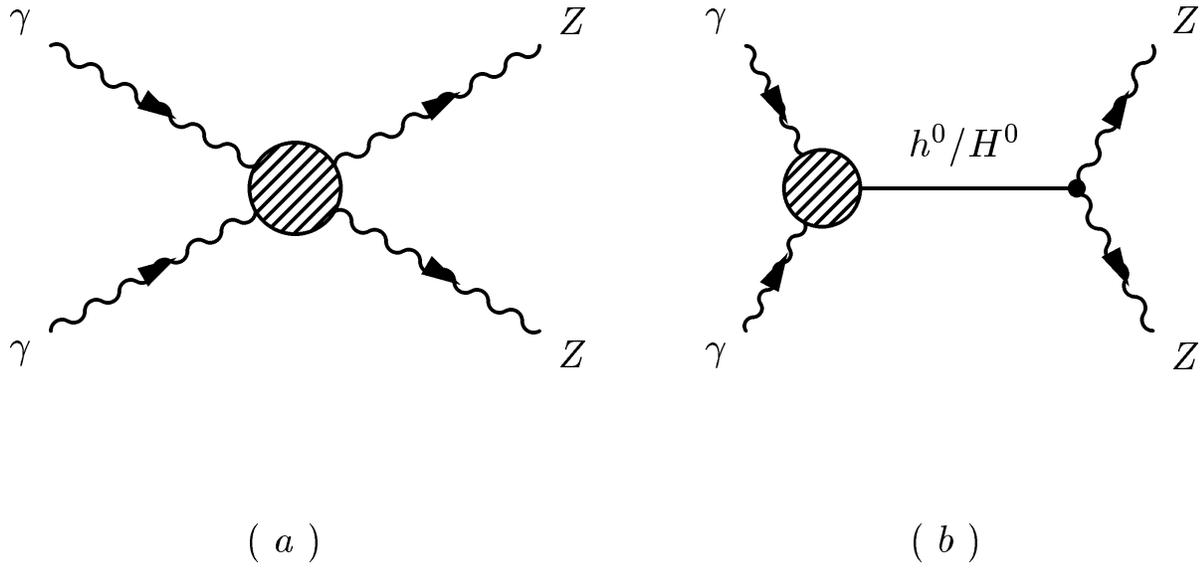,height=7.5cm}
\]
\caption[1]{Feynman Diagams for the $\gamma \gamma \to ZZ$
process in SM and MSSM models.}
\label{diag-fig}
\end{figure}

\begin{figure}[p]
\vspace*{-5.0cm}
\[
\hspace{-0.6cm}\epsfig{file=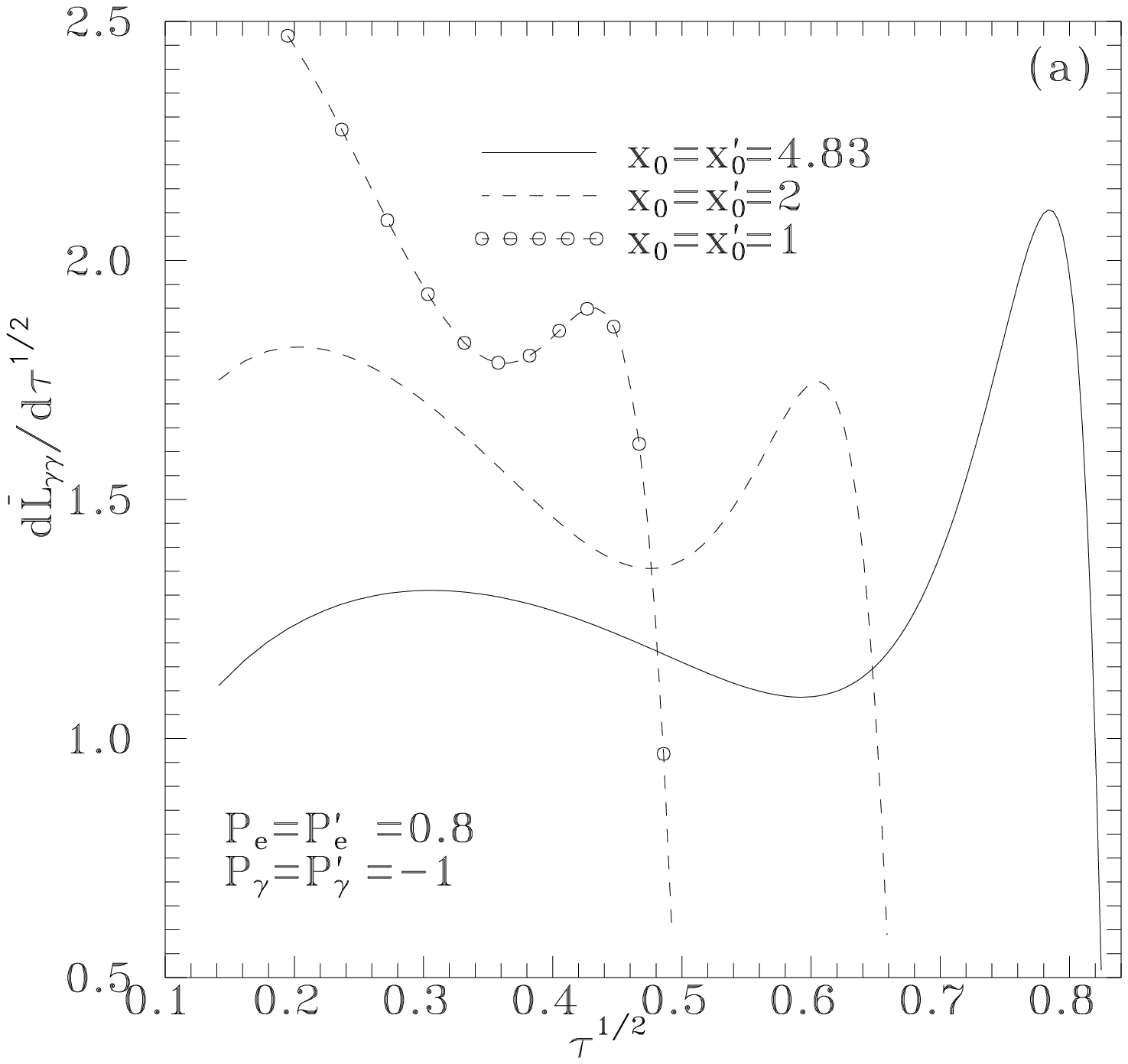,height=7.5cm}\hspace{0.5cm}
\epsfig{file=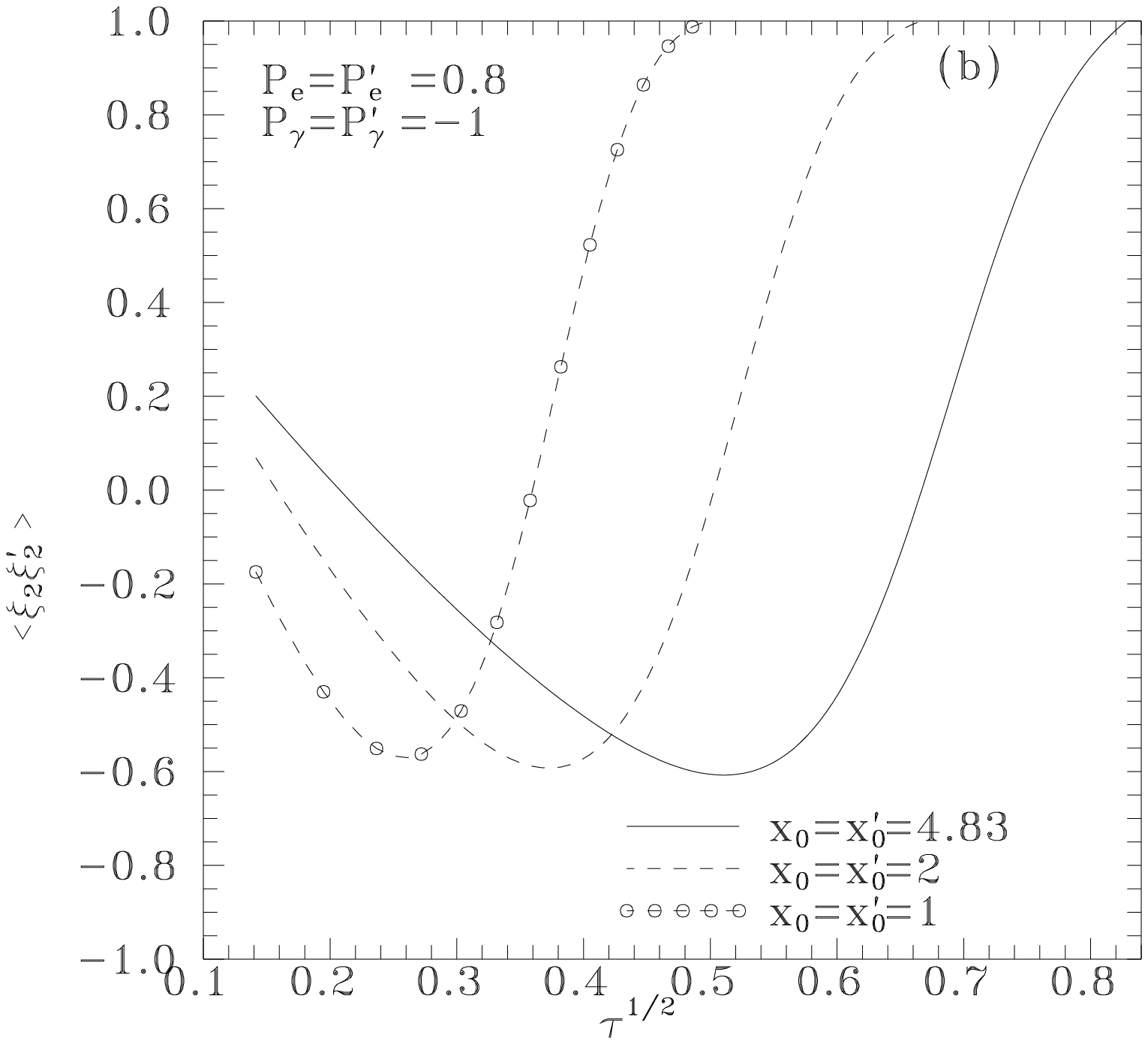,height=7.5cm}
\]
\vspace*{0.5cm}
\[
\epsfig{file=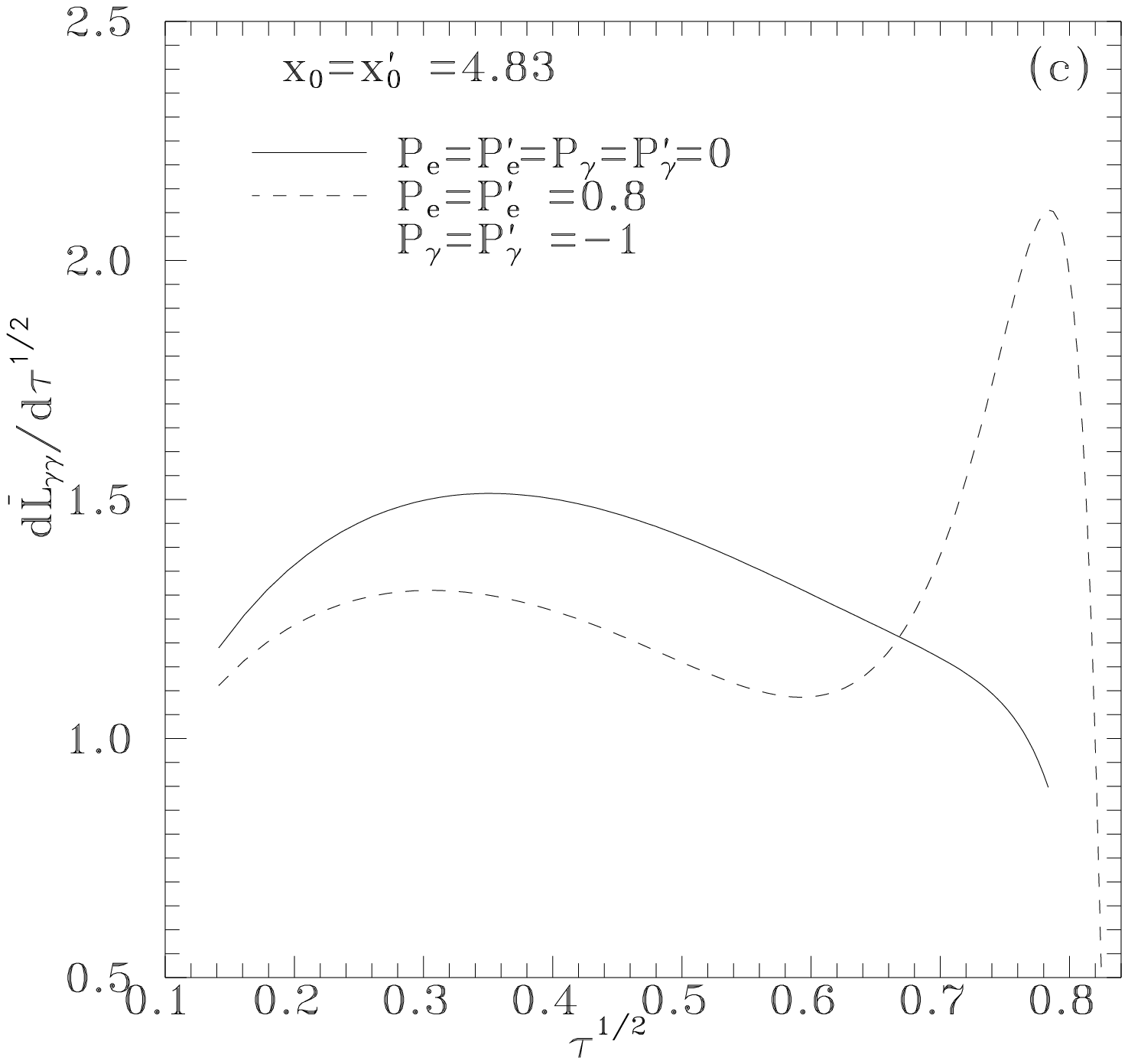,height=7cm}
\]
%
\caption[1]{Photon-photon luminosity factor (a), and circular
polarization factor (b), for  longitudinally  polarized
$e^{\pm}$ beams and  circularly polarized laser photons;
while (c) gives the same  luminosity factor for unpolarized
$e^{\pm}$ beams and laser photons.
The laser parameters $x_0,~ x_0^\prime$,
are indicated in the figures.}
\label{spectra}
\end{figure}

\begin{figure}[t]
\vspace*{-2cm}
\[
\epsfig{file=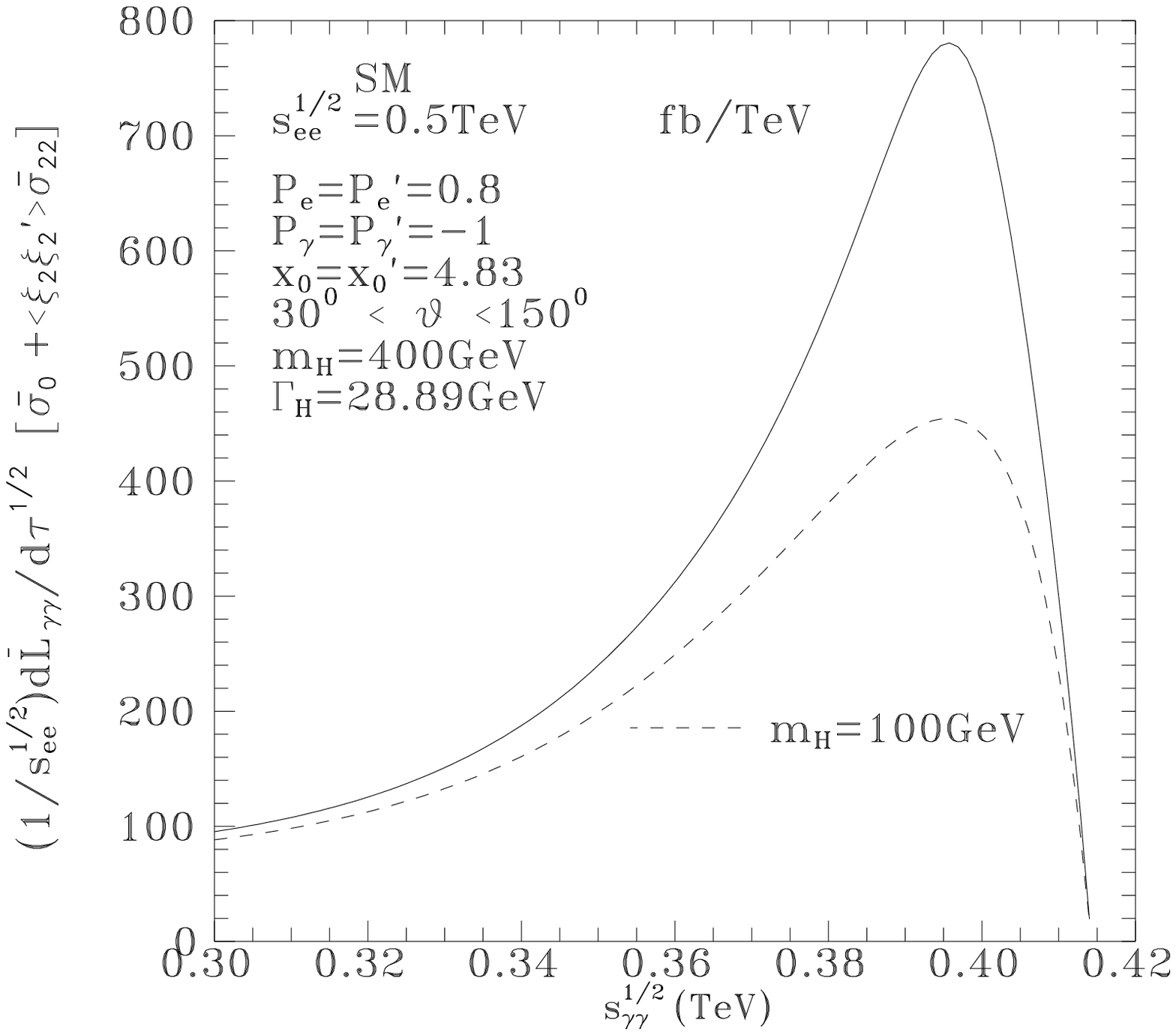,height=7.5cm}
\]
\caption[1]{A 0.5 TeV Linear Collider picture of the  SM contribution
to the $\sigma(\gamma \gamma \to Z Z)_{\rm Laser}$
cross section for a standard Higgs with $m_H=400~\rm GeV$.
 The dash lines give the results for
$ m_H =100~\rm GeV$. The machine is  assumed to run at 0.5TeV
total $e^-e^+$ energy
using the polarizations and $x_0$, $x_0^\prime$ values
indicated in the figure.}
\label{SM-4-fig}
\end{figure}

\begin{figure}[p]
\vspace*{-5cm}
\[
\hspace{-0.6cm}\epsfig{file=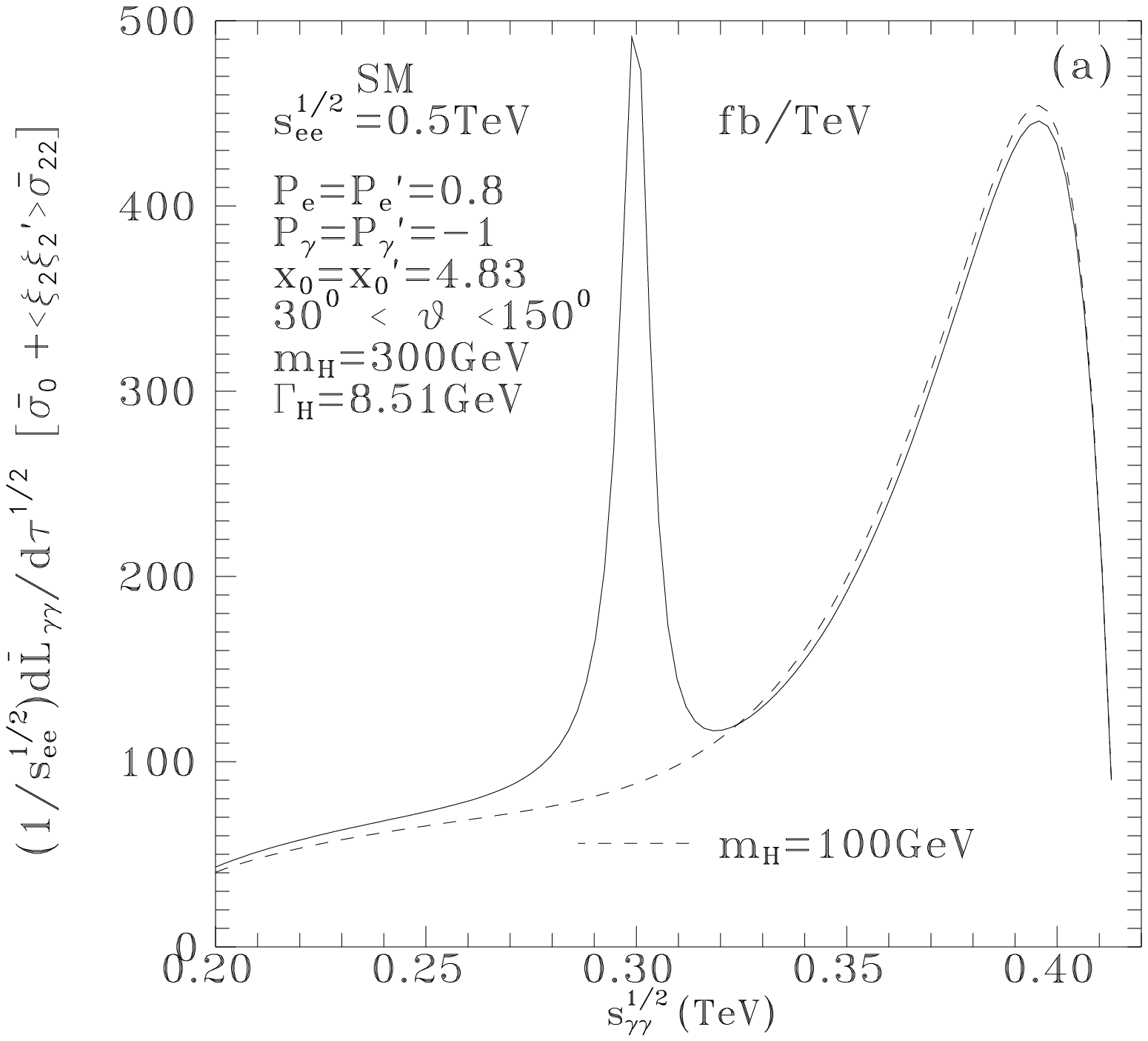,height=7cm}\hspace{0.5cm}
\epsfig{file=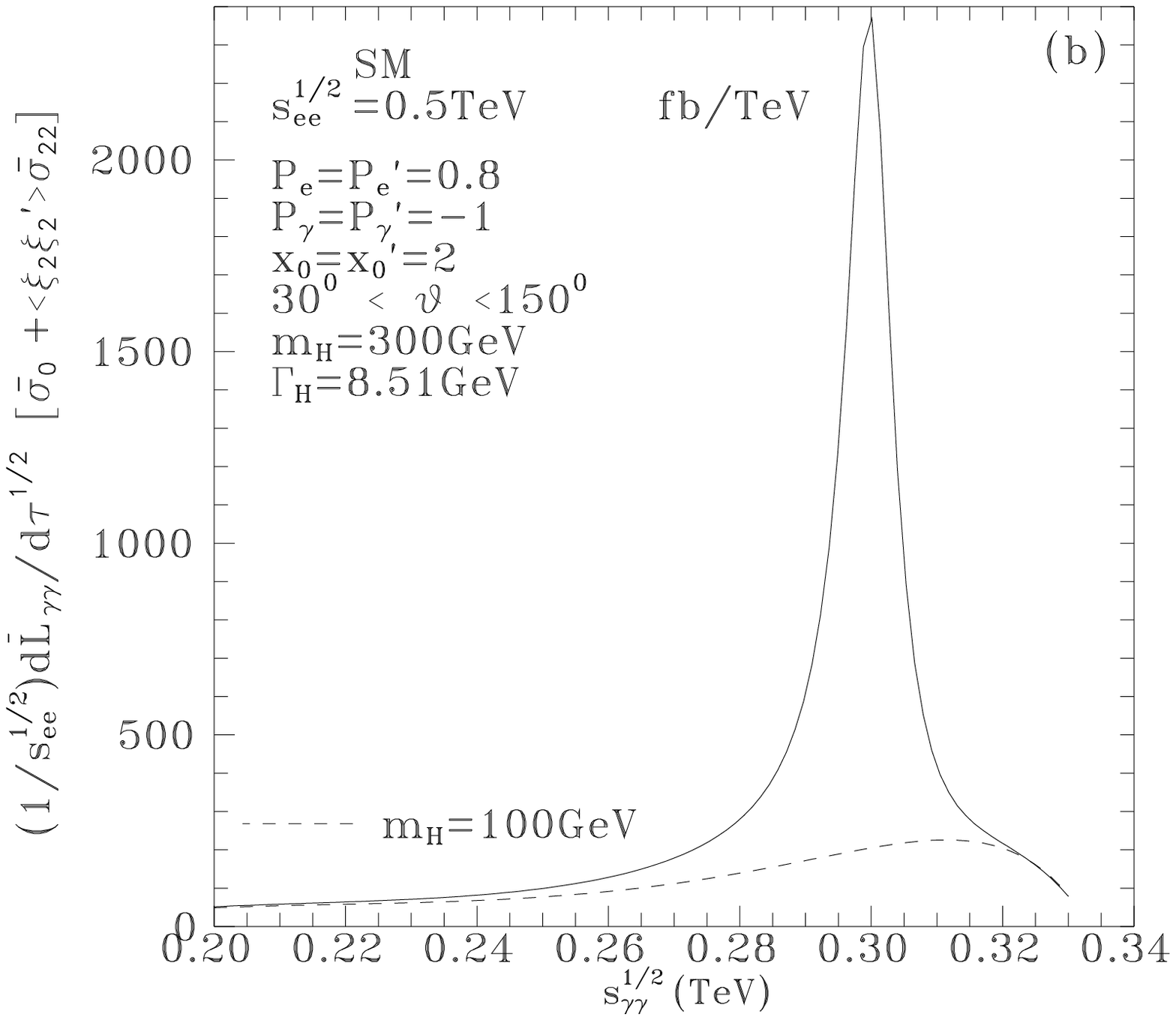,height=7cm}
\]
\vspace*{0.5cm}
\[
\hspace{-0.6cm}\epsfig{file=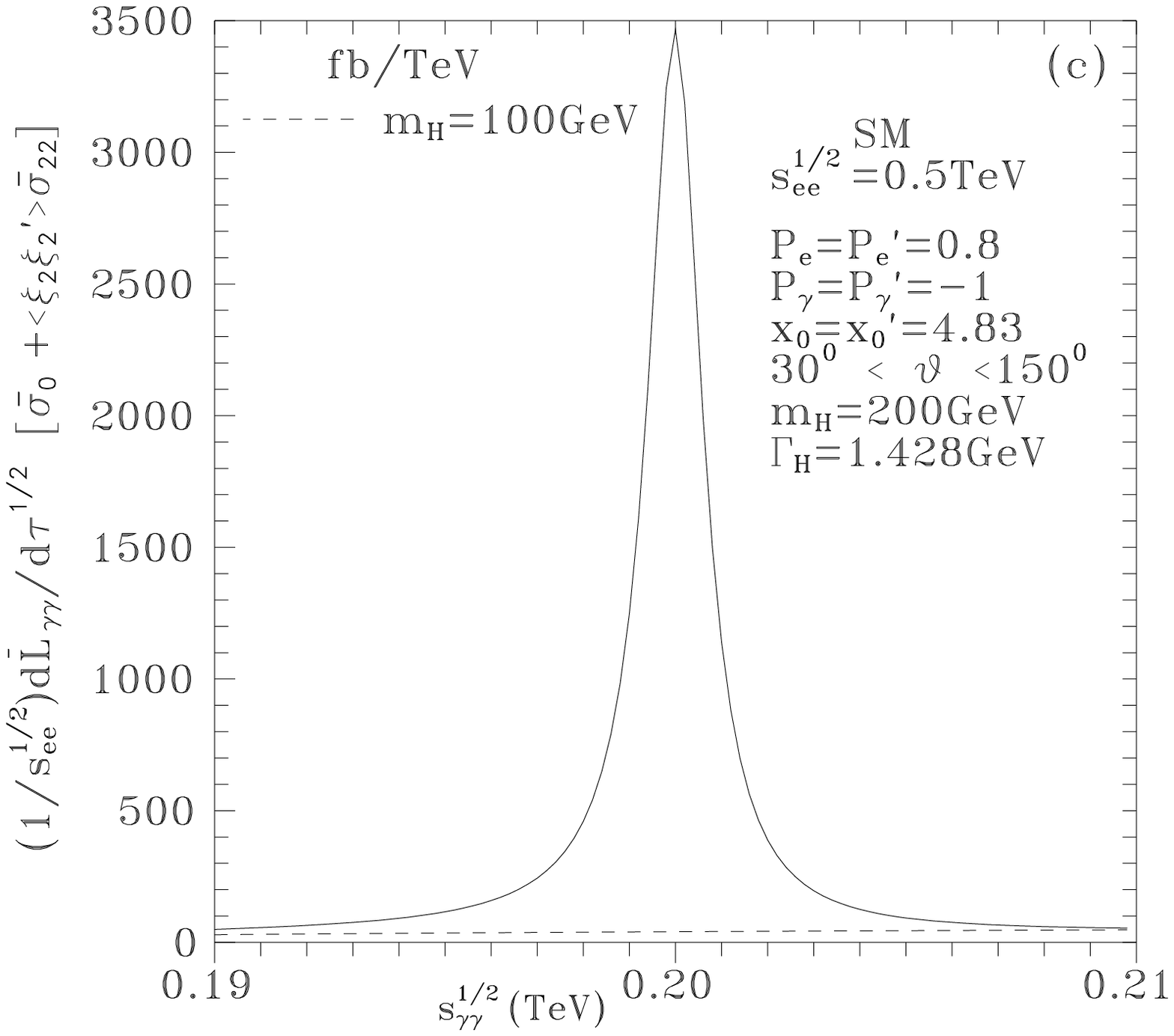,height=7cm}\hspace{0.5cm}
\epsfig{file=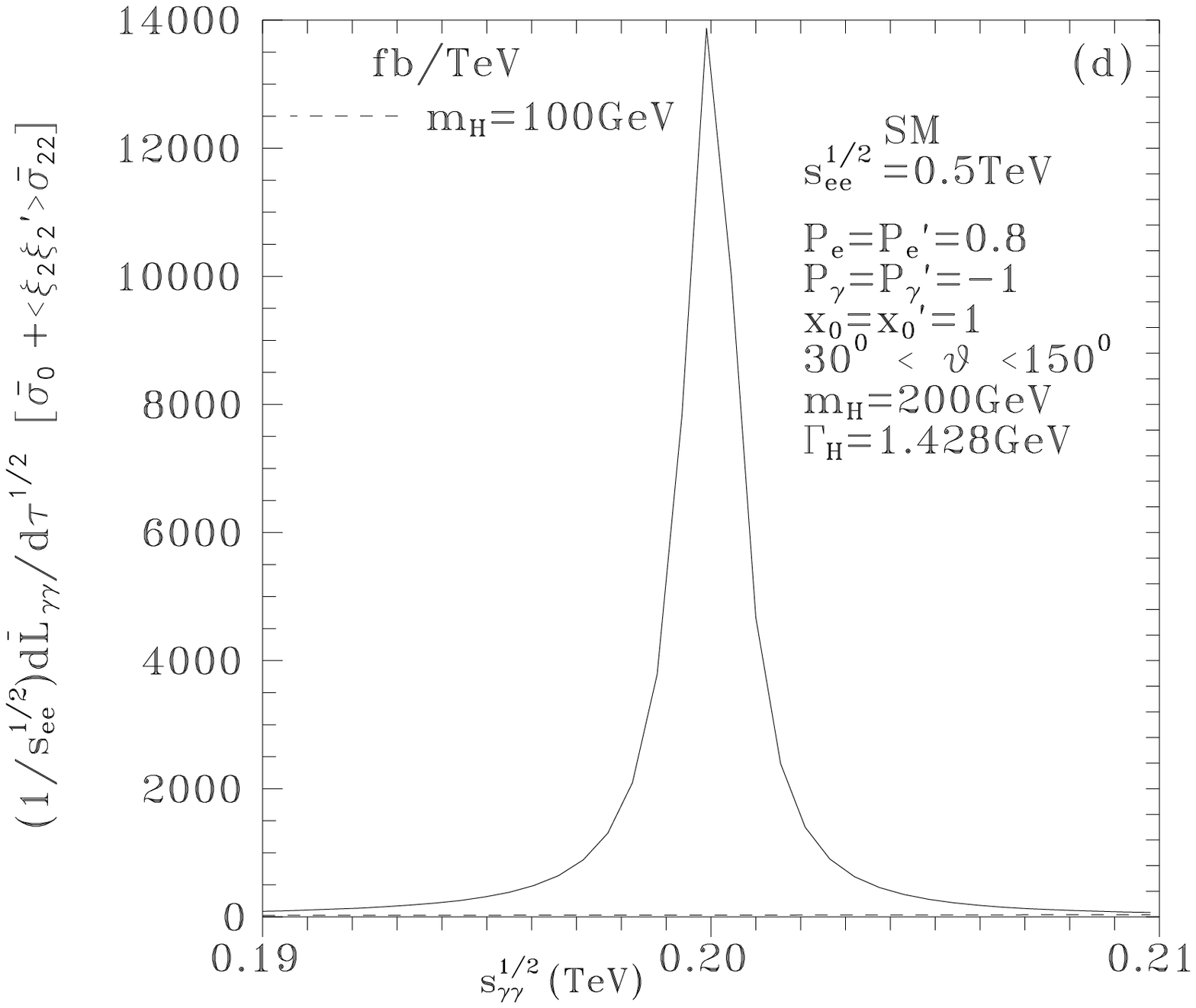,height=7cm}
\]
\caption[1]{A 0.5 TeV Linear Collider picture of the
SM contribution to the
$\sigma(\gamma \gamma \to Z Z)_{\rm Laser}$
cross section for a standard Higgs with $m_H=300~\rm GeV$ (a, b) and
$m_H= 200~\rm GeV$ (c, d). The dash lines give the results for
$ m_H =100~\rm GeV$. The machine is  assumed to run
at 0.5TeV total $e^-e^+$ energy
using the polarizations and $x_0$, $x_0^\prime$
values indicated in the figures.}
\label{SM-3-2-fig}
\end{figure}

\begin{figure}[p]
\vspace*{-5cm}
\[
\hspace{-0.6cm}\epsfig{file=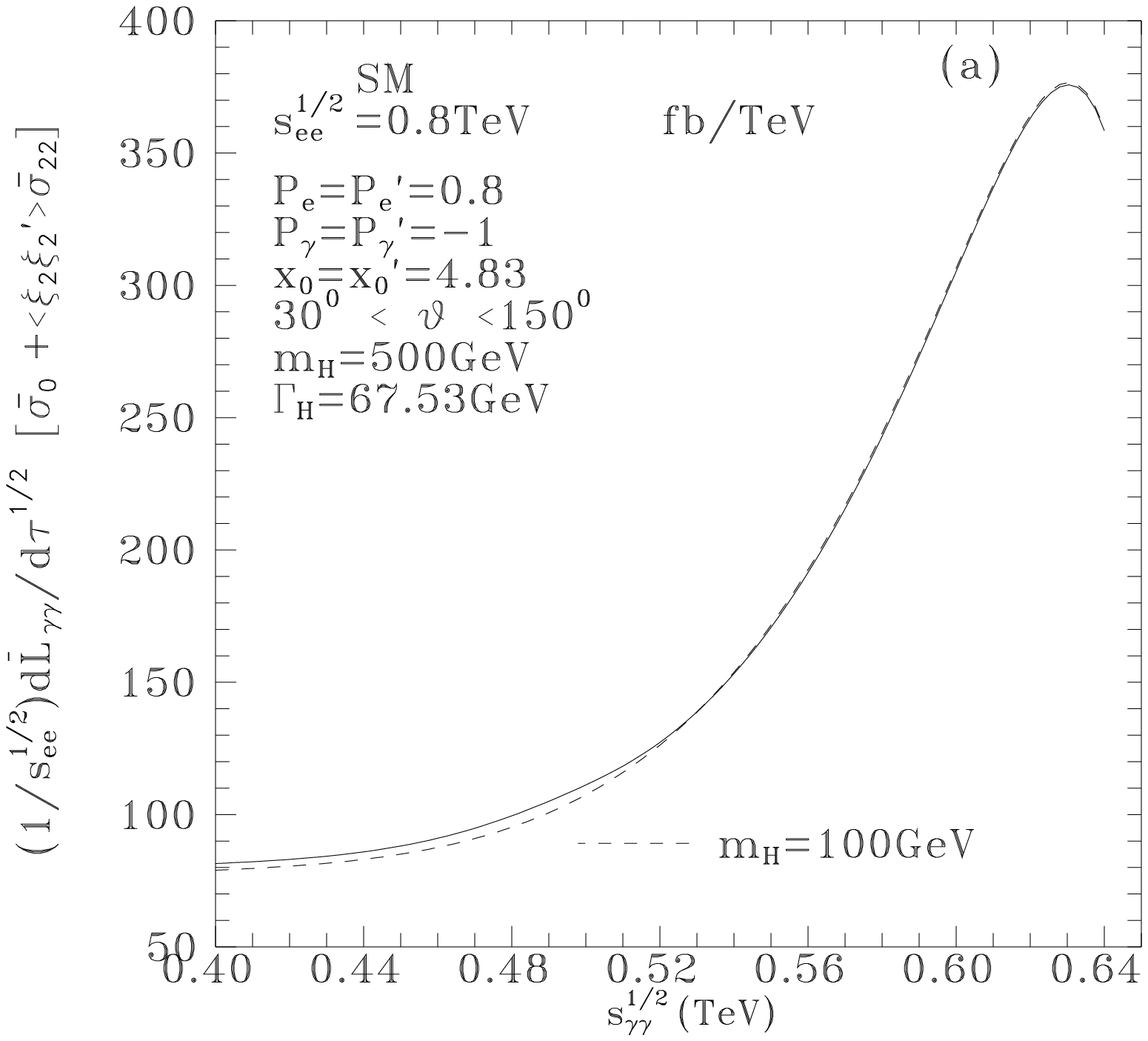,height=7cm}\hspace{0.5cm}
\epsfig{file=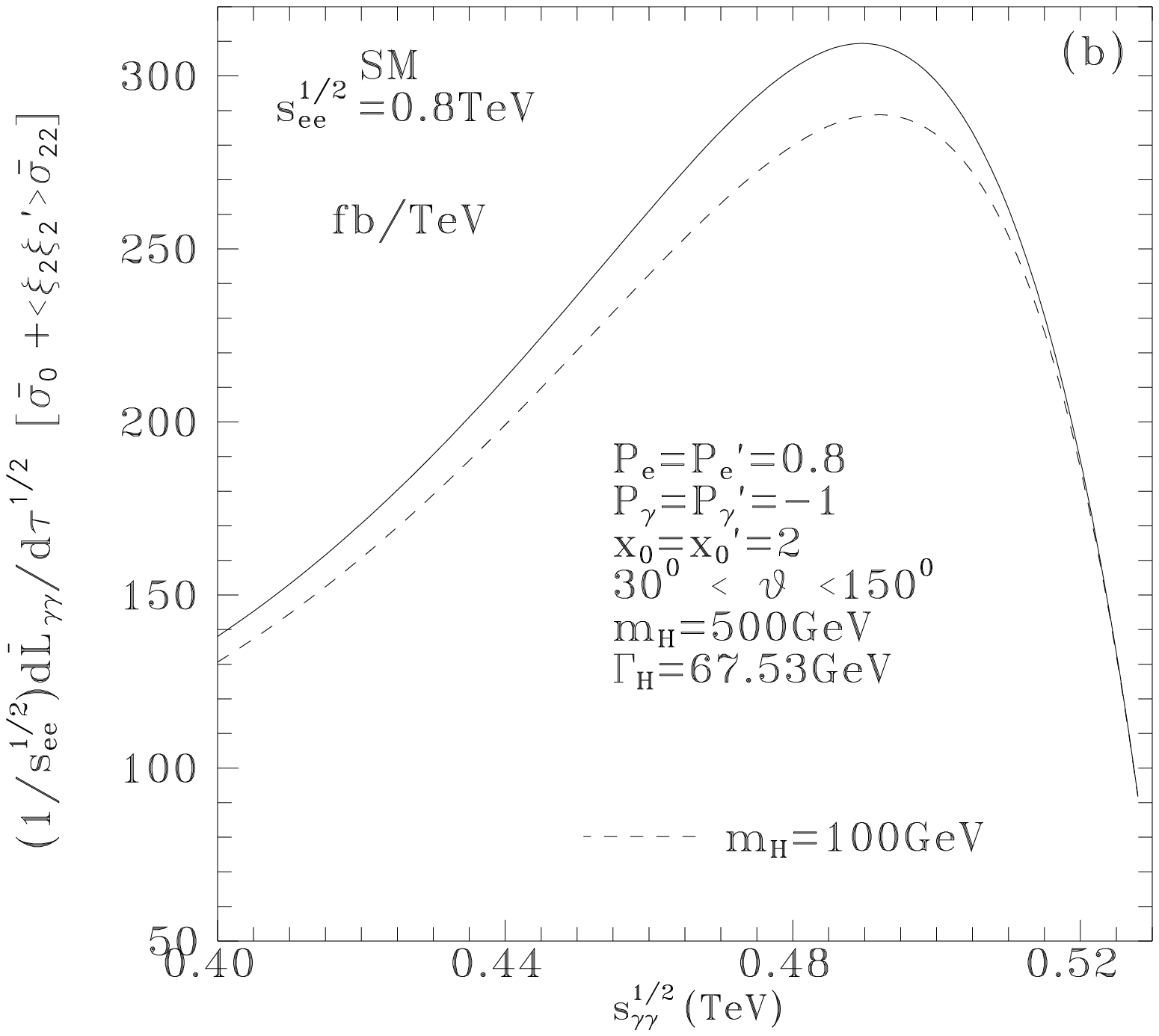,height=7cm}
\]
\vspace*{0.5cm}
\[
\epsfig{file=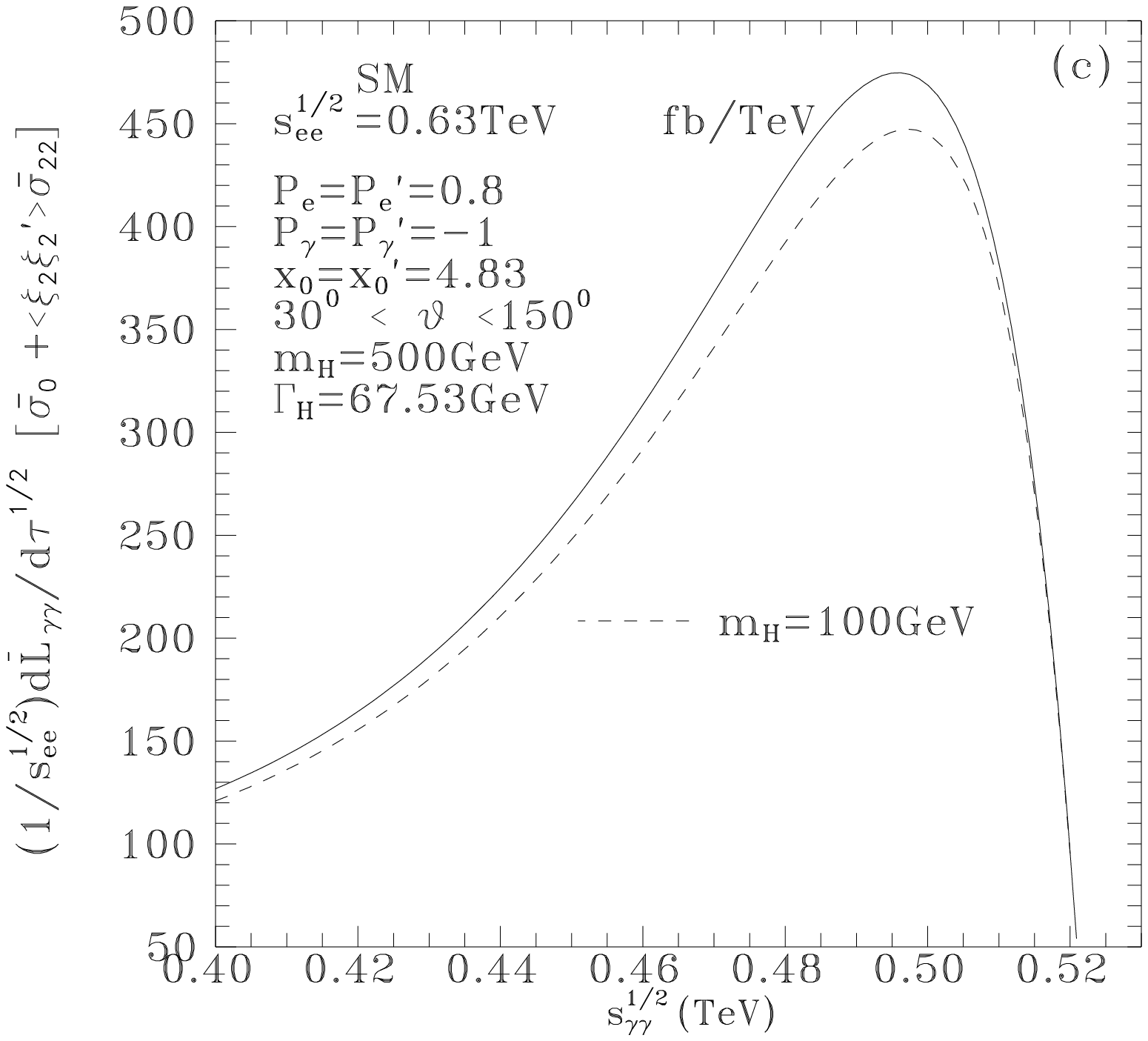,height=7cm}
\]
\caption[1]{A 0.8 TeV Linear Collider picture of the
SM contribution to the
$\sigma(\gamma \gamma \to Z Z)_{\rm Laser}$
cross section for a standard Higgs with $m_H=500~\rm GeV$.
The dash lines give the results for
$ m_H =100~\rm GeV$. In (a), (b) the machine is  assumed to run at
0.8TeV total $e^-e^+$ energy using the indicated  polarizations and
$x_0$, $x_0^\prime$ values; while in (c) the machine is tuned at
a total $e^-e^+$ energy of 0.63TeV .}
\label{SM-5-fig}
\end{figure}

\begin{figure}[p]
\vspace*{1.0cm}
\[
\hspace{-0.6cm}\epsfig{file=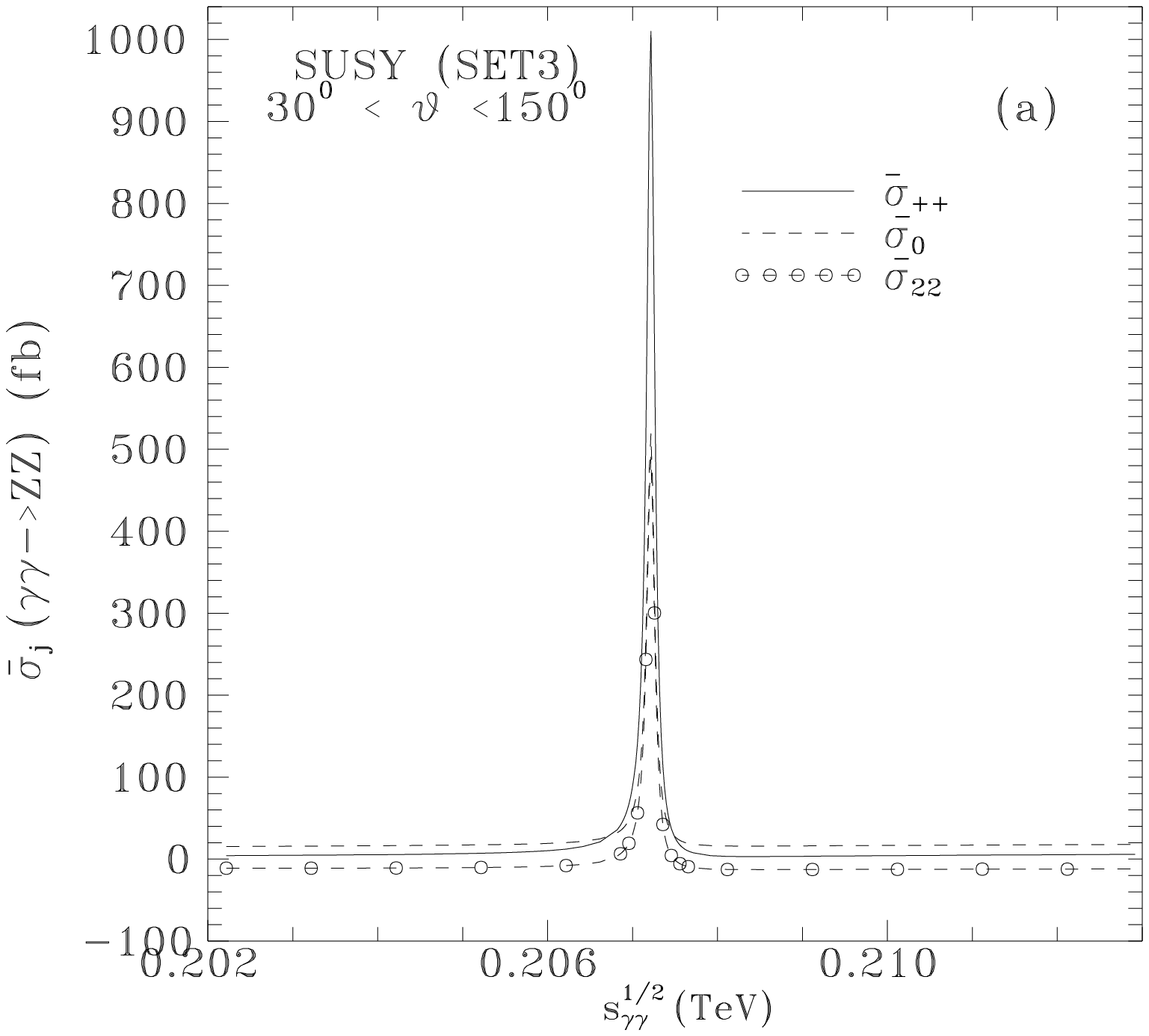,height=7cm}\hspace{0.5cm}
\epsfig{file=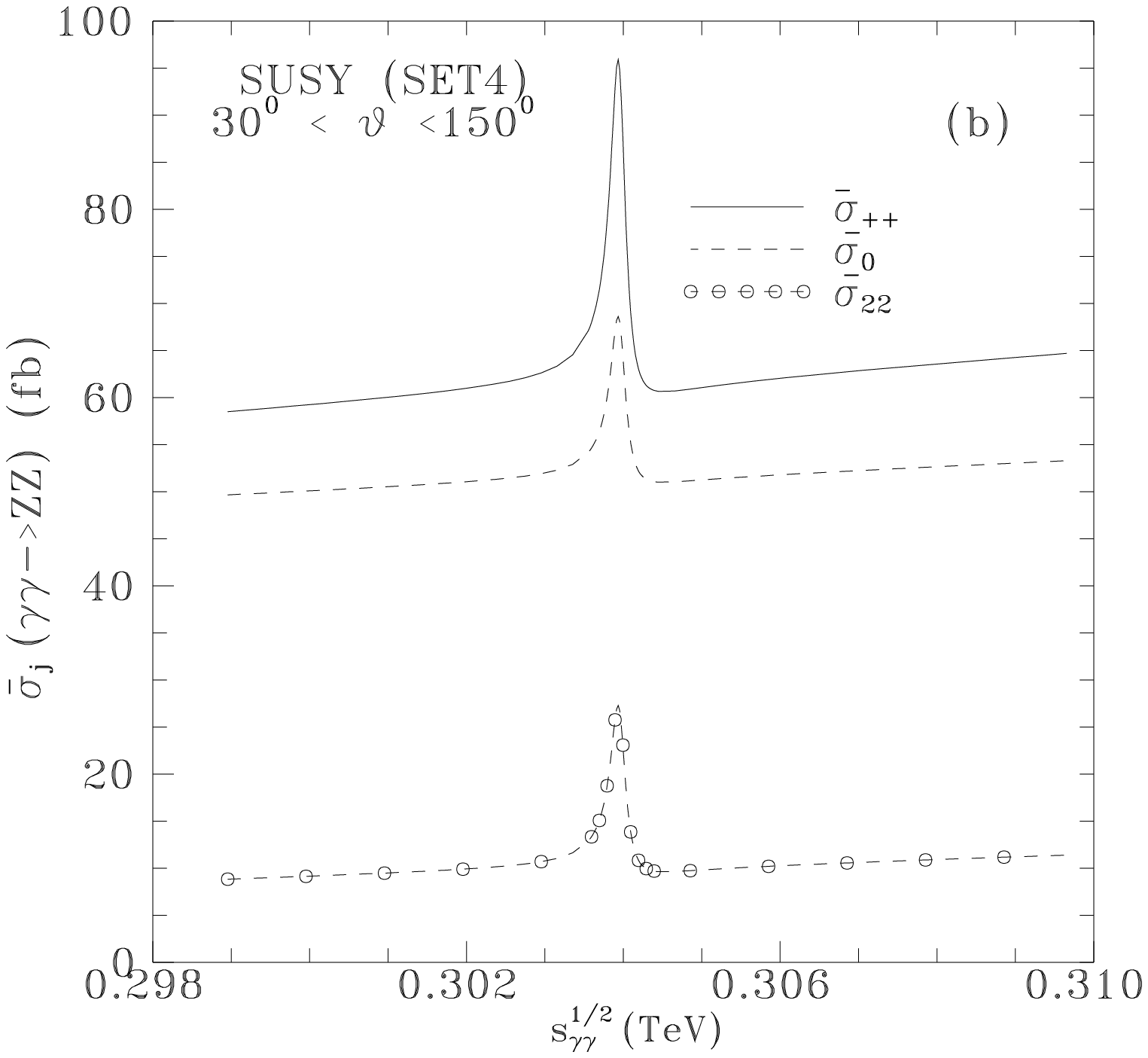,height=7cm}
\]
%
\caption[1]{SUSY predictions for the
cross sections $\bar \sigma_{0}(\gamma \gamma \to ZZ)$,
$\bar \sigma_{22}(\gamma \gamma \to ZZ)$ and
$\bar \sigma_{++}(\gamma \gamma \to ZZ)$ in  the $H^0$ mass region,
 using   the  parameters of
  Set3 (a) and  Set4 (b) in Table 2; see text.}
\label{SUSY-fig}
\end{figure}

\vspace{0.5cm}
In the non-linear gauge of \cite{Dicus} that we are using here,
there are only two types of contributions
to these amplitudes, in either the SM or SUSY models;
see Fig.\ref{diag-fig}.
The first  consists of the one-particle irreducible
 one-loop  diagrams involving four external legs,
similar to  those  contributing to the
$\gamma \gamma \to \gamma \gamma$ and
$\gamma \gamma \to \gamma Z$ processes
\cite{gggg, gggZ, Jikia-ZZ}.
We depict the generic form of these diagrams in
Fig.\ref{diag-fig}a  and call them "boxes".
Their contributions are given in this Appendix.
The second type (discussed in Section 2)
are one-particle reducible diagrams containing a
Higgs s-channel pole and involving an\footnote{Here $h^0$ denotes
any  neutral Higgs boson.}  $h^0 \gamma \gamma$
vertex subdiagram   \cite{Jikia-ZZ}; see Fig\ref{diag-fig}b. \par

\vspace{0.3cm}
\noindent
\underline{\bf The scalar boxes.} \\
Such contributions are generated in MSSM through the effective
Lagrangian
\bq
\L_{V \bar {S } S}  =
- i e (Q_{S} A^\mu +g^Z_{S} Z^\mu) ({S} ^* \partrl_\mu S)
+e^2 (Q_{S} A^\mu +g^Z_{S} Z^\mu)^2 |S |^2 ~~ ,
\label{gauge-scalar-vertex}
\eq
where $S$ is any scalar field. In the minimal SUSY case
where $S =\Stop_L, ~
\Stop_R$,  $\tilde b_L,~  \tilde b_R$, $ \tilde \tau_L, ~
\tilde \tau_R$ , $\tilde \nu_L$, or
\footnote{For $H^+$ we have $t_3^{H^+}=1/2$ and
$Q_{H^+}=1$.}  $ H^+ $, the corresponding
coupling is
\bq
g^Z_{S} = \frac{1}{\sw\cw} (t_3^{S} -Q_{S} \swd) ~,
\label{gZ-scalar}
\eq
in which $t_3^S$  denotes the third isospin component of $S$.

The contribution to $\gamma \gamma \to ZZ$ of any such scalar
particle is \cite{ggZZ}
\bq
F^S_{\lambda_1 \lambda_2 \lambda_3\lambda_4}(\beta_Z,\t,\u)\equiv
\alpha^2 Q^2_S N^c_S  \left (g^Z_{S}\right )^2
A^S_{\lambda_1 \lambda_2 \lambda_3\lambda_4} (\beta_Z,\t,\u; m)
~ , \label{F-scalar-amp}
\eq
where $N^c_{S}$ counts the
colour multiplicity of $S$, and
 $A^S_{\lambda_1 \lambda_2 \lambda_3\lambda_4}$ is given by
(A.34-A.41) in  \cite{ggZZ}.\par

In cases like $\Stop_{1,2}$, the non-diagonal mass matrix
\begin{equation}
    \left ( \matrix{\Stop_L \cr \Stop_R} \right ) =
    \left (
\matrix{\cos\theta_t &  - \sin\theta_t \Sn(A_t-\mu \cot \beta) \cr
\sin\theta_t\Sn(A_t-\mu \cot \beta)   & \cos\theta_t} \right )
    \left ( \matrix{\Stop_1 \cr \Stop_2} \right )
~ , \label{stop-angle-matrix}
\end{equation}
implies that the  mixing  angle always  satisfies
\bq
\frac{\pi}{2} < \theta_t < \pi ~~~~~~,\label{stop-angle-range}
\eq
and it is fully determined by
\bq
\sin(2\theta_t)=
\frac{2 m_t |A_t-\mu \cot\beta|}{m^2_{\Stop_1}-m^2_{\Stop_2}}
~~~~ ,  \label{stop-angle}
\eq
provided we define  $m_{\Stop_1} <m_{\Stop_2}$,  and $A_t$ is real.

Then,  the single $\Stop_1$-box contribution
is given by (\ref{F-scalar-amp}) for
\bq
g^Z_{\Stop_1} = \frac{1}{\sw\cw}\left [
\frac{1}{2}\cos^2 \theta_t -\frac{2\swd}{3} \right ] ~,
\label{gZ-stop1}
\eq
while for the single $\Stop_2$ one
\bq
g^Z_{\Stop_2} = \frac{1}{\sw\cw}\left [
\frac{1}{2}\sin^2 \theta_t -\frac{2\swd}{3} \right ] ~
\label{gZ-stop2}
\eq
should be used. In principle,  we should also
consider the mixed  box contribution arising
  when both $\Stop_1$ and $\Stop_2$ are
running along the box sides. Since such mixed contributions are
expected to be at most of similar magnitude to the  one coming from the
single $\Stop_1$-box \cite{neutral-model},
which is already known to be extremely small \cite{ggZZ}, we have not
calculated them.\par

If  $\tan\beta \gsim 10$, then  the $\tilde b_1$-squark or
$\tilde \tau_1$-slepton  contributions may be of similar magnitude.
If desired, they may be  directly obtained from
(\ref{F-scalar-amp}) using the appropriate mixing matrix.
 Since in the numerical examples we consider
these (as well as $\Stop_2$) are very heavy, we refrain from
giving their explicit contributions.

\vspace{0.3cm}
\noindent
\underline{\bf The $W$-boxes.}\\
These are 1-loop diagrams involving four external legs, with
a $W$, Goldstone  or FP-ghost running along the loop.
They have first been presented by \cite{Jikia-ZZ}.
We write them as
\bq
F^W_{\lambda_1 \lambda_2 \lambda_3\lambda_4}(\beta_Z,\t,\u)\equiv
\frac{\alpha^2}{\swd}
A^W_{\lambda_1 \lambda_2 \lambda_3\lambda_4} (\beta_Z,\t,\u)
~ , \label{F-W-amp}
\eq
with $A^W_{\lambda_1 \lambda_2 \lambda_3\lambda_4}$ given in
(A.42-A.51) of \cite{ggZZ}.\par

\vspace{0.3cm}
\noindent
\underline{\bf The fermion boxes.}\\
If the effective $(\gamma,~ Z)f\bar f$ interaction is written as
\bq
\L_{Vff} = -e Q_f A^\mu \bar f\gamma_\mu f  - e Z^\mu \bar f
(\gamma_\mu g_{vf}^Z- \gamma_\mu \gamma_5 g_{af}^Z) f ~~ ,
\label{gauge-fermion-vertex}
\eq
then the fermion loop contribution
to the $\gamma \gamma \to ZZ$ helicity amplitude
for a fermion of mass $m_f$,
is given by     \cite{Glover-ZZ, ggZZ}
\bqa
&& F^f_{\lambda_1 \lambda_2 \lambda_3\lambda_4}(\beta_Z,\t,\u)
\equiv \nonumber \\
&&
\alpha^2 Q^2_f N_f^c \Bigg \{ (g^Z_{vf})^2
A^{vf}_{\lambda_1 \lambda_2 \lambda_3\lambda_4} (\beta_Z,\t,\u; m_f )
+(g^Z_{af})^2
A^{af}_{\lambda_1 \lambda_2 \lambda_3\lambda_4} (\beta_Z,\t,\u; m_f)
\Bigg \} \ ,
\label{F-f-amp}
\eqa
where $N_f^c$ counts the colour multiplicity and
$A^{vf}$, $A^{af}$ are
 given by (A.55-A.71) of \cite{ggZZ}.

For quarks and leptons
\bq
g_{vf}^Z=\frac{t_3^f-2Q_f\sw^2}{2\sw\cw} ~~~~~ ,
~~~~~ g_{af}^Z=\frac{t_3^f}{2\sw\cw} ~ ~~~~ ,
\label{gZ-f}
\eq
where $t_3^f$ is the  third isospin component of the fermion,
and  $Q_f$ is its charge.

The specific  case of a  \underline{chargino} fermion
requires a more extensive  discussion, because of
their possible mixed coupling to $Z$.
The relevant parameters are determined by the mass matrix
\bq
\L_{M_\chi} =-
\left ( \matrix{ \tilde W^{-\tau} , \tilde H_1^{-\tau}}
\right )_L  \cdot C \cdot \left (\matrix{
  M_2 & \sqrt{2} \mw \sin\beta \cr
 \sqrt{2} \mw \cos \beta   &  + \mu \cr } \right )
 \left (\matrix {\tilde W^+ \cr \tilde H_2^+}\right )_L
~+ {\rm h.c.} ~   , \label{chi-mass-matrix}
\eq
leading to the physical chargino masses
\bq
m_{\tilde \chi_1, \tilde \chi_2}
=\frac{1}{\sqrt{2}} [M_2^2 +\mu^2 +2\mwd \mp \tilde D ]^{1/2}
~ , \label{chi-mass}
\eq
where
\bq
\tilde D \equiv  \left [
(M_2^2+\mu^2+ 2\mwd)^2- 4 (M_2\mu-\mwd \sin(2\beta))^2
\right ]^{1/2} ~ , \label{Dtilde}
\eq
for any sign of $M_2,~ \mu$.
Defining then the mixing-angles $\phi_R, \phi_L$  as
 \cite{ggAA}
\bqa
\cos\phi_L &=& -~ \frac{1}{\sqrt{2\tilde D}}
[\tilde D-M_2^2+\mu^2 +2\mwd \cos 2\beta ]^{1/2} ~~ ,
\nonumber \\
\cos\phi_R &=& -~ \frac{1}{\sqrt{2\tilde D}}
[\tilde D-M_2^2+\mu^2 -2\mwd \cos 2\beta ]^{1/2} ~~ ,
\label{chi-angles}
\eqa
 so that they always lie in the second quarter
\bq
\frac{\pi}{2} \leq \phi_L < \pi  ~~~~ , ~~~~
\frac{\pi}{2} \leq \phi_R  < \pi  ~~~ ; \label{chi-angle-range}
\eq
 the effective Lagrangian for the  $(\gamma, ~Z)$-chargino
interaction
becomes\footnote{The chargino field  is always defined so
that it absorbs a  positive chargino particle; \ie
$\tchi_j \equiv {\tchi_j}^+$ $(j=1,2)$.}
\bqa
\L & = & -e A^\mu \bar{\tchi}_j \gamma_\mu \tchi_j
-e   Z^\mu \bar{\tchi}_j \left  ( \gamma_\mu g_{vj}
- \gamma_\mu \gamma_5 g_{aj} \right ) \tchi_j
\nonumber \\
&& -e  Z^\mu \left [\bar{\tchi}_1 \left
( \gamma_\mu g_{v12}
- \gamma_\mu \gamma_5 g_{a12} \right ) \tchi_2 +
\mbox{h.c.} \right ]
  ~~ , \label{gauge-chi-vertex}
\eqa
where
\bqa
g_{v1}&= &\frac{1}{2\sw\cw}\left (
{3\over2}-2s^2_W+{1\over4}[\cos2\phi_L+\cos2\phi_R]\right ) ~ ,
 \nonumber \\
g_{a1}& = & - ~\frac{1}{8\sw\cw}[\cos2\phi_L-\cos2\phi_R] ~ ,
\label{gZ-chi1} \\
g_{v2}& = & \frac{1}{2\sw\cw}\left ({3\over2}-2s^2_W-~
{1\over4}[\cos2\phi_L+\cos2\phi_R] \right )
~ , \nonumber \\
g_{a2}& = & \frac{1}{8\sw\cw}[\cos2\phi_L-\cos2\phi_R] ~ ,
\label{gZ-chi2} \\
g_{v12}& = &
-~{\Sn(M_2)\over 8\sw\cw }
[\B_R ~ \Del_{12} \sin2\phi_R+ \B_L \sin2\phi_L] ~ ,
\nonumber\\
g_{a12}&= &-~{\Sn(M_2)\over 8\sw\cw}
[\B_R ~  \Del_{12}\sin2\phi_R - \B_L \sin2\phi_L] ~ .
\label{gZ-chi12}
\eqa\par

The box-contribution  from the single chargino
couplings in (\ref{gZ-chi1}, \ref{gZ-chi2}) are given by
the same expressions  (\ref{F-f-amp}).
But for charginos we
also have the "mixed" $Z \tchi_1\tchi_2$-couplings appearing  in
(\ref{gZ-chi12}), which generate boxes with two different
charginos running along the loop. These couplings,
as well as those of the neutral Higgs to charginos defined
in Section 2,
depend on the sign-quantities
\bqa
\B_L & = &\Sn (\mu \sin\beta +M_2 \cos\beta)  ~,
\nonumber \\
\B_R & = & \Sn (\mu \cos\beta +M_2 \sin \beta) ~,
\nonumber \\
\B_{LR} & \equiv & \Sn \left(M_2 \mu+
\frac{\mu^2+M_2^2}{2}\sin 2\beta \right ) =\B_L \B_R ~ ,
\nonumber \\
\Del_1 &= &
\Sn (M_2 [\tilde D-M_2^2+\mu^2-2\mwd] -2 \mwd \mu \sin 2\beta )  ~, ~
\nonumber \\
 \Del_2 &=&
\Sn (\mu [\tilde D-M_2^2+\mu^2 +2\mwd] +2 \mwd M_2 \sin 2\beta ) ~ ,
\nonumber \\
\Del_{12} &\equiv & \Sn (M_2 \mu -\mwd \sin 2\beta)=
 \Del_1  \Del_2 ~ , \label{signs}
\eqa
constructed to guarantee the positivity of the physical chargino
masses and the usual
 relation between the fields absorbing the  positive and negative
charginos; \ie $C \bar{\tchi}^{+\tau}=\tchi^-$ \cite{ggAA}. \par

\vspace{0.3cm}
\noindent
\underline{\bf The mixed chargino  boxes.}\\
 This contribution,
generated by the $Z \tchi_1\tchi_2$ - couplings in
(\ref{gZ-chi12}), is denoted as\footnote{The factor
$(-1)^{1-\lambda_4}$ comes from the JW helicity convention
\cite{JW-convention}.}
\bqa
&& F^{\tchi_1 \tchi_2}_{\lambda_1 \lambda_2
\lambda_3\lambda_4}(\beta_Z,\t,\u)
\equiv
\alpha^2   [(g_{v12})^2+ (g_{a12})^2](-1)^{1-\lambda_4}
A^{(\tchi_1\tchi_2 1)}_{\lambda_1 \lambda_2 \lambda_3\lambda_4}
(\beta_Z,\t,\u; m_{\tchi_1^2}, m_{\tchi_2^2})
\nonumber \\
&& + \alpha^2   [(g_{v12})^2- (g_{a12})^2]
m_{\tchi_1} m_{\tchi_2}~(-1)^{1-\lambda_4}
A^{(\tchi_1\tchi_2 2)}_{\lambda_1 \lambda_2 \lambda_3\lambda_4}
(\beta_Z,\t,\u; m_{\tchi_1^2}, m_{\tchi_2^2}) \ .
\label{F-chi1chi2-amp}
\eqa
The form of (\ref{F-chi1chi2-amp})
is motivated by the fact that the structure of the mixed boxes
allow only the existence of either $g_{v12}^2$ or $g_{a12}^2$
terms, which are  related  to each other through
the substitutions:
\[
g_{v12} \leftrightarrow g_{a12}
~~~~~~~~{\rm and} ~~~~~~~~
(m_1 ~,~ m_2)  \leftrightarrow  (- m_1 ~, ~  m_2) ~~ .
\]
To describe this mixed chargino contribution, we need
 the  Passarino-Veltman functions
\cite{Passarino}, for which we follow the
notation of \cite{Hagiwara} and the
abbreviations\footnote{In the middle terms
of (\ref{C0s}-\ref{DZZut}) $k_1=p_1$, $k_2=p_2$ denote
the momenta of the photons, while  $k_3=-p_3$, $k_4=-p_4$
those of the $A^0$, always taken as incoming;
compare (\ref{ggZZ-process}).}
\bqa
&& B_Z^{11}(\s) \equiv   B_0(\s; m_1, m_1)
-B_0(m^2_Z +i\epsilon; m_1, m_2) \ , \label{BZ11} \\
&& B_Z^{22}(\s) \equiv   B_0(\s; m_2, m_2)
-B_0(m^2_Z +i\epsilon; m_1, m_2) \ , \label{BZ22} \\
&& B_Z^{12}(\s) \equiv   B_0(\s; m_1, m_2)
-B_0(m^2_Z +i\epsilon; m_1, m_2) \ , \label{BZ12} \\
&& C_{0}^{abc}(\s)\equiv
 C_0(k_1, k_2) =  C_{0}(0,0,\s; m_{a},m_{b},m_{c}) ~ ,
\label{C0s} \\
&& C_{Z}^{abc}(\u)\equiv
C_0(k_3, k_2) =  C_{0}(m_{Z}^2,0,\u; m_{a},m_{b},m_{c})
~ ,\label{CZu} \\
&& C_{ZZ}^{abc}(\s) \equiv C_0(k_3, k_4)  =
C_{0}(m_{Z}^2,m_{Z}^2,\s;m_{a},m_{b},m_{c})
~ , \label{CZZs} \\
&& D_{ZZ}^{abcd}(\s,\t) \equiv D_0(k_4, k_3, k_1) =
D_{0}(m_{Z}^2,m_{Z}^2,0,0,\s,\t;m_{a},m_{b},m_{c},m_{d})
~ , \label{DZZst}  \\
&& D_{ZZ}^{abcd}(\s,\u) \equiv D_0(k_3, k_4, k_1) =
D_{0}(m_{Z}^2,m_{Z}^2,0,0,\s,\u;m_{a},m_{b},m_{c},m_{d})
~ , \label{DZZsu}  \\
&& D_{ZZ}^{abcd}(\t,\u)\equiv D_0(k_3, k_1, k_4)
 = D_{0}(m_{Z}^2,0,m_{Z}^2,0,\t,\u; m_{a},m_{b},m_{c},m_{d})
~ , \label{DZZtu} \\
&& D_{ZZ}^{abcd}(\u,\t)\equiv D_0(k_4, k_1, k_3)
 = D_{0}(m_{Z}^2,0,m_{Z}^2,0,\u,\t; m_{a},m_{b},m_{c},m_{d})
~ , \label{DZZut}
\eqa
\bqa
\tilde F^{ab}(\s,\t,\u) &= &D_{ZZ}^{abba}(\t,\u)
+D_{ZZ}^{abaa}(\s,\t)+
D_{ZZ}^{abaa}(\s,\u) ~ , \label{Fstu} \\
E_{1}^{ab}(\s,\u)& = & 2 \u_1 C_{Z}^{baa}(\u)-
\s \u D_{ZZ}^{abaa}(\s,\u) ~ , \label{E1st} \\
E_{2}^{ab}(\t,\u)& = &\t_1 \left [ C_{Z}^{abb}(\t)+C_{Z}^{baa}(\t)
\right ]+ \u_1 \left [ C_{Z}^{abb}(\u)+C_{Z}^{baa}(\u) \right ]
\nonumber \\
&- &Y D_{ZZ}^{abba}(\t,\u) ~, \label{E2tu}
\eqa
which are closely related to  those in\footnote{Notice
that the present definition of $E_1$ differs somewhat from the
one employed in \cite{ggAA}, where an analogous mixed case is
also treated.}  Eqs.(A.14 - A.24) of \cite{ggZZ}.

We also  note that
\bqa
& D_{ZZ}^{abba}(\t,\u) = D_{ZZ}^{abba}(\u,\t) =
D_{ZZ}^{baab}(\t,\u)= D_{ZZ}^{baab}(\u,\t)
~~ , &~~  \nonumber  \\
& \tilde F^{ab}(\s,\t,\u)=\tilde F^{ab}(\s,\u,\t)
~~~, ~~~ E_{2}^{ab}(\t,\u) =E_{2}^{ab}(\u,\t) =
 E_{2}^{ba}(\t,\u)~~ . &
\label{D-E2-relations}
\eqa \par

Thus,  the eight basic amplitudes listed
in (\ref{8basic}, \ref{F-chi1chi2-amp}) are determined
by\footnote{For brevity we identify here
$m_j\equiv m_{\tchi_j}$.}
\bqa
&& A^{(\tchi_1\tchi_2 1)}_{++++}
(\beta_Z,\t,\u; m_1^2, m_2^2)
= -\frac{16[\mzd (2Y-\s\s_4)+\beta_Z \s Y]}{\s_4 \t_1\u_1}
\nonumber \\
&& +\frac{4 (\s_2+\s\beta_Z)}{\s_4\s} \Bigg [
\frac{ (2 Y -\s\s_4)}{\s}E_2^{12}(\t,\u)
+\frac{4\mzd Y+2 \t_1(2\t_1+\s)(\t+\mzd)}{\t^2_1} B_Z^{12}(\t)
\nonumber \\
&&+ \frac{4\mzd Y+2\u_1(2\u_1+\s)(\u+\mzd)}{\u^2_1} B_Z^{12}(\u)
\Bigg ] \nonumber \\
&& +\frac{8}{\s_4} \Bigg \{-2 m_1^2 [2
(\mzd+m_2^2-m_1^2)+\s\beta_Z]C_0^{111}(\s)+
\frac{(m_1^2-m_2^2)^2}{\s}[E_2^{12}(\t,\u)
\nonumber \\
&& -2 \s m_1^2 \tilde F^{12}(\s,\t,\u)]
-\frac{(m_1^2+m_2^2)(2Y-\s\s_4)(\s_2+\beta_Z \s)}
{4\s}D_{ZZ}^{1221}(\t,\u)
\nonumber \\
&& -2m_1^2 \mzd(\s_2+\beta_Z \s)\left [\frac{1}{\u_1}C_Z^{211}(\u)+
\frac{1}{\t_1}C_Z^{211}(\t)\right ] +m_1^2 [\s (m_1^2-m_2^2)
\nonumber \\
&& -\s_2 \mzd -\beta_Z\s (\mzd+m_2^2-m_1^2)][D_{ZZ}^{1211}(\s,\u)
+ D_{ZZ}^{1211}(\s, \t)] + ( 1 \leftrightarrow 2) \Bigg \} ~ ,
\label{Amix1++++}
\eqa
\bqa
&& A^{(\tchi_1\tchi_2 2)}_{++++}
(\beta_Z,\t,\u; m_1^2, m_2^2) =\frac{8}{\s_4} \Bigg \{
2\beta_Z\s C_0^{111}(\s) +\frac{\s_4+\beta_Z\s}{\s}E_2^{12}(\t,\u)
-2 m_1^2\s_4 \tilde F^{12}(\s,\t,\u)
\nonumber \\
&& +\frac{\s}{2}[\s_4 +\beta_Z (s_2 + 2 m_2^2-2m_1^2)]
[D_{ZZ}^{1211}(\s,\u)+D_{ZZ}^{1211}(\s,\t)]+
( 1 \leftrightarrow 2) \Bigg \} ~ , \label{Amix2++++}
\eqa
\bqa
&& A^{(\tchi_1\tchi_2 1)}_{+-++}
(\beta_Z,\t,\u; m_1^2, m_2^2)
=\frac{16 \s_2 Y}{\s_4\t_1\u_1}-
\frac{4 \mzd \s(\s_2^2-2Y)}{\s_4 Y}[C_0^{111}(\s)+C_0^{222}(\s)]
\nonumber \\
&& +\frac{4\mzd \s_2(2Y-\s\s_4)}{\s_4 Y}[C_{ZZ}^{121}(\s)+
C_{ZZ}^{212}(\s)] -\frac{4\mzd}{\s_4} \Big \{
\frac{4 (Y+2\u \mzd)}{\u_1^2}B_Z^{12}(\u)
\nonumber \\
&& +\frac{(\u^2+\mz^4)}{Y}[E_1^{12}(\s,\u)+E_1^{21}(\s,\u)]
+ ( \u \leftrightarrow \t) \Big \}
\nonumber \\
&& +\frac{4}{\s_4 Y} \Bigg \{ \s(m_1^2-m_2^2)[4
(m_1^2-m_2^2)(\mzd +m_1^2-m_2^2-\s)
+\s_2(\s+2\mzd)-2Y]C_0^{111}(\s)
\nonumber \\
&& -(2Y-\s\s_4)(m_1^2-m_2^2)[\s-2 (m_1^2-m_2^2)]C_{ZZ}^{121}(\s)
-2\s(m_1^2-m_2^2)^4 \tilde F^{12}(\s,\t,\u)
\nonumber \\
&& - \left (2 Y(m_1^4-m_2^4)(m_1^2-m_2^2)-(2Y-\s\s_4)\mzd
\Big [(m_1^2-m_2^2)^2 +\frac{(m_1^2+m_2^2)Y}{\s}\Big ]
\right )D_{ZZ}^{1221}(\t,\u)
\nonumber \\
&& +\Big [ \Big ( \frac{2\u_1}{\s}(m_1^2-m_2^2)[\u_1
(2 \mz^4 -3\mzd \s +\u\s_2)-(m_1^2-m_2^2)\s(\s-2\u)]-
\frac{4m_1^2\s_2 Y^2}{\s\u_1} \Big )C_Z^{211}(\u)
\nonumber \\
&& - \Big ((m_1^6- m_2^6)\s(4\u-\s)- (m_1^2-m_2^2)\s(\mz^6
-\u\u_1^2+\mzd \t\u +2\mzd \u^2)
\nonumber \\
&& +m_1^2m_2^2 (m_1^2-m_2^2)(8 \u_1^2+3 \s^2 -4\s\u)
+(m_1^2-m_2^2)^2 \s(\s\mzd +4 \u\u_1)
\nonumber \\
&& -(m_1^4-m_2^4)\s Y
+2 Y m_1^2[2 (m_1^4-m_2^4)+\s_2 \mzd] \Big ) D_{ZZ}^{1211}(\s,\u)
+ ( \u \leftrightarrow \t) \Big ]
\nonumber \\ &&
+ ( 1 \leftrightarrow 2) \Bigg \}~, \label{Amix1+-++}
\eqa
\bqa
&& A^{(\tchi_1\tchi_2 2)}_{+-++}
(\beta_Z,\t,\u; m_1^2, m_2^2)
= \frac{8}{Y} \Bigg \{ \s[2(m_1^2-m_2^2)-\s_2]C_0^{111}(\s)
+(2Y-\s\s_4)C_{ZZ}^{121}(\s)
\nonumber \\
&& -\s(m_1^2-m_2^2)^2\tilde F^{12}(\s,\t,\u,)
-\frac{Y}{\s_4}[\s_4 (m_1^2+m_2^2)+Y]D_{ZZ}^{1221}(\t, \u)
\nonumber \\ &&
+\Big [ \u E_1^{12}(\s,\u)+2 (m_1^2\u_1^2+m_2^2\s\u)
D_{ZZ}^{1211}(\s,\u)+ ( \u \leftrightarrow \t) \Big ]
+ ( 1 \leftrightarrow 2) \Bigg \}~, \label{Amix2+-++}
\eqa
\bqa
&& A^{(\tchi_1\tchi_2 1)}_{+++-}
(\beta_Z,\t,\u; m_1^2, m_2^2)
= \frac{16 Y}{\s_4} \Big [ \frac{\s_2}{\t_1\u_1}
-\frac{\mzd}{\s^2} E_2^{12}(\t,\u)
+\frac{\mzd}{\t_1^2} \Big (\frac{2\t}{\s}-1 \Big )B_Z^{12}(\t)
\nonumber \\ &&
+\frac{\mzd}{\u_1^2}B_Z^{12}(\u) \Big ]
+\frac{8}{Y\s_4} \Bigg  ( -m_1^2[2(m_1^2-m_2^2)-\s_2]
(2Y-\s\s_4) C_0^{111}(\s)
-m_1^2\s\s_4^2 C_{ZZ}^{121}(\s)
\nonumber \\ &&
+ \frac{Y\s_4}{2\s}(m_1^2+m_2^2)E_2^{12}(\t,\u)
+m_1^2 (m_1^2-m_2^2)^2 (2Y-\s\s_4)\tilde F^{12}(\s,\t,\u)
\nonumber \\ &&
+\frac{Y}{\s} \Big [Y [\mzd (m_1^2+m_2^2)-2 m_1^2m_2^2]
+(m_1^4+m_2^4)(Y-\s\s_4) \Big ] D_{ZZ}^{1221}(\t,\u)
\nonumber \\ &&
+\Big \{ \frac{1}{\s\u_1} \Big [-\u_1^2 (m_1^2-m_2^2)^2
(2Y -\s\s_4) +2\s m_1^2 [2\mz^4 (Y +2\u^2)-\u^2\s_2^2]
\Big ] C_Z^{211}(\u)
\nonumber \\  &&
+ m_1^2 [-2 (m_1^2-m_2^2)(\mz^4 \t -3 \mz^4 \u - \u^2\s_2)
+2 Y (\mz^4- m_1^2 \s_4)
\nonumber \\ &&
+(4 \mz^4-\s_2^2 ) \u^2 ]
D_{ZZ}^{1211}(\s, \u) + ( \u \leftrightarrow \t) \Big \}
+ ( 1 \leftrightarrow 2) \Bigg  ) ~, \label{Amix1+++-}
\eqa
\bq
 A^{(\tchi_1\tchi_2 2)}_{+++-}
(\beta_Z,\t,\u; m_1^2, m_2^2)
= 0 ~ , \label{Amix2+++-}
\eq
\bqa
&& A^{(\tchi_1\tchi_2 1)}_{+-00}
(\beta_Z,\t,\u; m_1^2, m_2^2)
= -\frac{64 \mzd Y}{\s_4 \t_1 \u_1} +
\frac{8\s\mzd (\s_2^2-2Y)}{\s_4 Y} [C_0^{111}(\s)+C_0^{222}(\s)]
\nonumber \\ &&
-\frac{8 \mzd \s_2 (2Y-\s \s_4)}{\s_4 Y}
[C_{ZZ}^{121}(\s) +C_{ZZ}^{212}(\s)] +
\Big \{ \frac{8\mzd}{\s_4Y}(\u^2 +\mz^4)
[E_1^{12}(\s,\u)+E_1^{21}(\s,\u)]
\nonumber \\ &&
-\frac{32 \mzd (\u^2+\mz^4)}{\s_4 \u_1^2} B_Z^{12}(\u)
+ ( \u \leftrightarrow \t) \Big \}
\nonumber \\ &&
+\frac{2}{\s_4 Y \mzd}
\Bigg [ -\s [4\s (m_1^2-m_2^2)^3
-16 m_1^2 \mzd (m_1^2-m_2^2)(\s-\mzd)
\nonumber \\ &&
+ 4 m_2^2 (m_1^2-m_2^2)(4 \mz^4 +\s\s_2)
+ 4(m_1^2-m_2^2) \mzd (4 \mzd \s_2 +\s_2^2 -2Y)
\nonumber \\ &&
-(m_1^2+m_2^2) \s_4 (\s_2^2-2Y) ]C_0^{111}(\s)
-( 2 Y-\s\s_4) [(m_1^2+m_2^2) (\s^2+8 \mz^4)
+2 \s (m_1^2-m_2^2)^2
\nonumber \\ &&
 -2 \mzd \s (5 m_1^2+m_2^2)]
C_{ZZ}^{121}(\s)
+\frac{1}{\s} \Big \{ 2\s^3 (m_1^2-m_2^2)^4
+\s\s_2 (\s-8\mzd)Y(m_1^4+m_2^4)
\nonumber \\ &&
+(m_1^2 -m_2^2)(m_1^4-m_2^4)\s^2 (\s_4\s_2 +2Y)
- 4 \mzd \s_2 Y^2 (m_1^2+m_2^2)
+2 m_1^2 m_2^2 \s^2 \s_2 Y \Big \} D_{ZZ}^{1221}(\t,\u)
\nonumber \\ &&
+ \Bigg \{ \frac{2}{\s\u_1}
\Big [ - 2(m_1^2-m_2^2)^2 (\u- 2\mzd)\u_1^2 \s^2
+8 m_1^2 \s\mzd (\mz^6 \t +5 \mz^6 \u -3 \mz^4 \t\u
\nonumber \\ &&
-7\mz^4 \u^2 +\mzd \t \u^2 +5 \mzd \u^3 -\t\u^3 -\u^4)
+ (m_1^2+m_2^2)\u_1^2 [8 \u_1^2 \mz^4 + 5 \mz^4\s\s_2
\nonumber \\ &&
- 2 \mzd \s (2 \u\s_2 +\mz^4)+\s\s_4 \u^2 ] \Big ]
C_Z^{211}(\u) + \Big \{
2 \s^2 (m_1^2-m_2^2)^4 +m_2^6 \s(\s^2-2 \mzd \s_4-4 \s\u)
\nonumber \\ &&
-m_1^4 m_2^2 \s (\s^2- 18\s\mzd +16\u \mzd +4 \s\u -8 \u^2)
+m_1^6 \s [ 8 \mz^4 + \s(\s-10\mzd)-4 \u_1^2]
\nonumber \\ &&
- m_1^2 m_2^4 \s (28 \mz^4 -10 \mzd \t +\t^2 -34 \mzd \u
+10\t\u +13 \u^2)
-(m_1^2-m_2^2)^2\s (8 \mz^6
\nonumber \\ &&
-5 \mz^4 \s +16\mz^4 \u +\s\s_2\u -3\s\u^2 )
-4 m_1^4 [ 4 \u_1^2 \mz^4 +\s (\u^2 +\mz^4)(\s- 7\mzd)
\nonumber \\ &&
+\s\u\mzd (10\mzd -3 \s)]- 4 m_1^2 m_2^2
[\mzd (\s+4 \mzd)(\u^2 +\mz^4)-8 \mz^6\u - 5 \mzd \s\s_2\u
\nonumber \\ &&
+\s^3\u ]- m_2^2 \s (-4 \mz^8- 8 \mz^6 \u +5 \mz^4 \s\u
-12 \mz^4 \u^2 +\s\u^3 )
\nonumber \\ &&
+m_1^2 (16 \mz^6 \u_1^2 -12 \mz^8 \s
+32 \mz^6 \s\u -5\mz^4 \s^2 \u -20 \mz^4 \s\u^2
+8 \mzd \s\u^3
\nonumber \\ &&
 -\s^2\u^3 ) \Big \} D_{ZZ}^{1211}(\s,\u)
+( \u \leftrightarrow \t) \Bigg \}
+( 1 \leftrightarrow 2) \Bigg ] ~ , \label{Amix1+-00}
\eqa
\bqa
&& A^{(\tchi_1\tchi_2 2)}_{+-00}
(\beta_Z,\t,\u; m_1^2, m_2^2)
= \frac{2}{\s_4 Y \mzd} \Bigg [
2\s_4 \s [2 (m_1^2-m_2^2)\s_2 -\s_2^2 +2 Y]C_0^{111}(\s)
\nonumber \\ &&
-2 (6 \mzd \s -8\mz^4 -\s^2)(2Y -\s\s_4)C_{ZZ}^{121}(\s)
-2[ \s\s_2\s_4 (m_1^2-m_2^2)^2+\s_4\s_2 Y(m_1^2+m_2^2)
\nonumber \\ &&
-4 \mzd Y^2 ] D_{ZZ}^{1221}(\t,\u) -2\s_4 \Big \{
(\u^2+\mz^4)E_1^{12}(\s,\u) + [ (m_1^2-m_2^2)^2 \s\s_2
+m_2^2 \s (2\u^2 +\t\u +\mz^4)
\nonumber \\ &&
+m_1^2(4 \mzd \u_1^2 -3\s\u_1^2 -\s^2\u)]D_{ZZ}^{1211}(\s,\u)
+( \u \leftrightarrow \t) \Big \}
+( 1 \leftrightarrow 2) \Bigg ] ~ , \label{Amix2+-00}
\eqa
\bqa
&& A^{(\tchi_1\tchi_2 1)}_{+++0}
(\beta_Z,\t,\u; m_1^2, m_2^2)
= \frac{8 (1+\beta_Z)Y}{\s_4} \Big [
-\frac{2(\t-\u)}{\t_1\u_1}-
\frac{2(\mzd \s -2 \u\u_1)}{\s\u_1^2} B_Z^{12}(\u)
\nonumber \\ &&
+\frac{2 (\mzd \s -2 \t\t_1)}{\s\t_1^2} B_Z^{12}(\t)
+\frac{(\t-\u)}{\s^2} E_2^{12}(\t ,\u) \Big ]
+\frac{2}{\s_4\mzd} \Bigg [
2 m_1^2 (\t-\u)[4 (m_1^2-m_2^2)
\nonumber \\ &&
-(1+\beta_Z)\s]C_0^{111}(\s)
-\frac{(\t-\u)}{2\s}  \Big \{ (m_1^2-m_2^2)^2
[4\s (m_1^2+m_2^2) -(\t-\u)^2 ]
\nonumber \\ &&
+ 4 \mzd (m_1^2+m_2^2) Y (1+\beta_Z)
+ 4 \beta_Z m_1^2 m_2^2 \s\s_4
-\beta_Z (m_1^2+m_2^2)^2 \s\s_4 \Big \} D_{ZZ}^{1221}(\t, \u)
\nonumber \\ &&
+ \frac{\s_4 (\u-\t)(m_1^2+m_2^2) (1+\beta_Z)}{2\s}
E_2^{12}(\t,\u)  +\Big \{
\frac{4}{\s\u_1} [-\u_1^2 (\u-\t) (m_1^2-m_2^2)^2
\nonumber \\ &&
+2 (1+\beta_Z)m_1^2 \mzd \s (\mz^4 - Y -\u^2)]C_Z^{211}(\u)
-m_1^2 [4 (m_1^2-m_2^2)
\nonumber \\ &&
-(1+\beta_Z)\s][(m_1^2-m_2^2)(\t-\u) +\mz^4-Y -\u^2]
D_{ZZ}^{1211}(\s, \u)
- ( \u \leftrightarrow \t) \Big \}
\nonumber \\ &&
+( 1 \leftrightarrow 2) \Bigg ]
~ , \label{Amix1+++0}
\eqa
\bqa
&& A^{(\tchi_1\tchi_2 2)}_{+++0}
(\beta_Z,\t,\u; m_1^2, m_2^2)
= \frac{2(\s_4 +\beta_Z \s)}{\s_4 \mzd} \Bigg [
2 (\t-\u) C_0^{111}(\s) +\frac{(\t-\u)}{\s} E_2^{12}(\t, \u)
\nonumber \\ &&
+ [(m_1^2-m_2^2)(\u-\t) +\mz^4 -Y-\t^2]
D_{ZZ}^{1211}(\s,\t)
\nonumber \\ &&
- [(m_1^2-m_2^2)(\t-\u) +\mz^4 -Y-\u^2]
D_{ZZ}^{1211}(\s,\u)
+( 1 \leftrightarrow 2) \Bigg ] ~ , \label{Amix2+++0}
\eqa
\bqa
&& A^{(\tchi_1\tchi_2 1)}_{+-+0}
(\beta_Z,\t,\u; m_1^2, m_2^2)
= - \frac{16 Y( \u-\t-\s\beta_Z)}{\s_4\t_1\u_1}+
\frac{8 (\u-\t+\s \beta_Z)}{\s_4}
[ B_Z^{11}(\s)+B_Z^{22}(\s)]
\nonumber \\ &&
- \frac{4\s}{\s_4 Y} [ (\t-\u)(Y + \u^2+\t^2)
+\beta_Z (\t\t_1^2 +\u\u_1^2 -2\mzd Y)]
[C_0^{111}(\s)+C_0^{222}(\s)]
\nonumber \\ &&
+\frac{4}{\s_4 Y} \{ (\u-\t)\s_4 (\t_1^2 +\u_1^2+Y)
-\beta_Z \s [ \t\t_1 (\t+\mzd)+\u\u_1(\u+ \mzd)
\nonumber \\ &&
- 2\mzd Y ]\} [C_{ZZ}^{121}(\s)+C_{ZZ}^{212}(\s)]
+\Big \{ \frac{16}{\s_4 \u_1^2}[\mzd Y
-\u\u_1 (\u+\mzd) -\beta_Z( \mzd Y -\u\u_1^2) ]B_Z^{12}(\u)
\nonumber \\ &&
-\frac{4}{\s_4 Y} [\mzd Y -\u\u_1 (\u+\mzd)
+\beta_Z (\u\u_1^2 -\mzd Y)][E_1^{12}(\s,\u)+E_1^{21}(\s,\u)]
\nonumber \\ &&
- ( \u \leftrightarrow \t ~~, ~~ \beta_Z \to -\beta_Z ) \Big \}
+ \frac{2}{\mzd \s_4 Y} \Bigg [
- 4 Y (m_1^2-m_2^2)(\u-\t+\beta_Z \s)B_Z^{11}(\s)
\nonumber \\ &&
+2 \{ (m_1^2-m_2^2) (\t-\u)
[ 2\s (m_1^2-m_2^2)(m_1^2-m_2^2+\mzd -\s) +
\s (Y+\t\t_1+\u\u_1 +\mzd \s_2)]
\nonumber \\ &&
+m_1^2 (\t-\u) \s_4 Y -\beta_Z \s [2\s (m_1^2-m_2^2)^3
+2 (m_1^4+m_2^4)(\mzd -\s) \s +4 m_1^4 Y
+4m_1^2m_2^2 (\t_1^2
\nonumber \\ &&
 +\u_1^2 +\mzd \s +Y) + (m_1^2-m_2^2)
(4 \mzd \u_1^2 +\s^2 \s_2 +2\s\u(\u+\s))-m_2^2 Y \s] \}
C_0^{111}(\s)
\nonumber \\ &&
+2 (m_1^2-m_2^2) \s (\s +2 m_2^2-2m_1^2 )
[\u(u+2\mzd)-\t(\t+2 \mzd)+\beta_Z( 2Y -\s\s_4)]C_{ZZ}^{121}(\s)
\nonumber \\ &&
-2 (m_1^2-m_2^2)^4 \s(\t-\u -\beta_Z \s)\tilde F^{12}(\s,\t,\u)
-\frac{Y}{2} \{(\t-\u)
[4 (m_1^2-m_2^2)(m_1^4-m_2^4)
\nonumber \\ &&
-\s(m_1^2-m_2^2)^2 -Y(m_1^2+m_2^2)] -\beta_Z
[8 \s (m_1^2-m_2^2)(m_1^4 -m_2^4)+(m_1^2+m_2^2)Y\s_4
\nonumber \\ &&
+(m_1^2-m_2^2)^2 (\s\s_4 +4Y)] \}D_{ZZ}^{1221}(\t,\u)
+\Bigg [ -\frac{1}{\u_1}
\Big \{ -4\u_1^2 (m_1^2-m_2^2)^2 (\s\mzd -2\u\u_1)
\nonumber \\ &&
+\u_1^2 (m_1^2 -m_2^2) (8 \mz^6 +\t Y -7\mz^4 \u -8\mzd \u^2
+3 \t\u^2 +4\u^3 ) +2m_1^2 (\u-\t)(\u+\mzd)^2 Y
\nonumber \\ &&
+ \beta_Z \{ 4(m_1^2-m_2^2)^2 \u_1^3 (\s-\u_1)
+(m_1^2-m_2^2) \u_1^2 [4 \mzd \u_1^2 -\s(\u+\mzd)^2
+\s\u (4\u-\s)]
\nonumber \\ &&
-2 m_1^2 Y [4 \mzd \u_1^2 +\s (\u+\mzd)^2 ] \} \Big
\} C_Z^{211}(\u) +\Big \{ -2 \s \u_1 (\s+4 \u)
(m_2^6-m_1^6 -m_1^4 m_2^2
\nonumber \\ &&
-m_1^2m_2^4)- 4 Y m_1^6 (\t-\u) -4 m_1^2m_2^2
(2m_1^2-m_2^2)(\s-2\u_1)(\mzd \t_1+\u\u_1)
\nonumber \\ &&
+(m_1^2-m_2^2)^2 \s (4 \mz^6 +\t Y -7 \mz^4 \u
+2 \mzd \t\u +6 \u^2 \u_1+\t\u^2)
-m_1^2 Y (m_1^2-m_2^2) ( 8\mzd \u_1
\nonumber \\ &&
+10 \mzd \s +2\s\u -\s^2)
-(m_1^2-m_2^2) \s (-2 \mz^4 Y -6 \mz^6 \u -\t\u Y
+3 \mz^4 \u^2+6 \mzd \u^3
\nonumber \\ &&
-\t\u^3 -2\u^4 ) +m_1^2 Y (8 \mz^6 -2 \mzd \t\u+\t^2 \u
-6 \mzd \u^2 -\u^3)
+\beta_Z \{ 2(m_2^6-m_1^6-m_1^4 m_2^2
\nonumber \\ &&
-m_2^4 m_1^2 )\s^2 (\mzd-3\u) +8\s Y m_1^6
-4 m_1^2 m_2^2 (2 m_1^2-m_2^2)\s\u_1 (\s-2\u_1)
\nonumber \\ &&
-2 (m_1^2-m_2^2)^2 \s\u_1(3\u_1^2-Y)
-m_2^2 (m_1^2-m_2^2)\s Y (6\mzd -\t -9\u)
\nonumber \\ &&
-2(m_1^2-m_2^2) \u_1^2 (-\mz^4 \s_4 -8\mz^4\u+3\mzd\s\u
-\s^2\u +4\mzd \u^2)
\nonumber \\ &&
+m_2^2 Y (8\mzd \u_1^2-\s\s_4 \u)
\} \Big \} D_{ZZ}^{1211}(\s,\u)
- ( \u \leftrightarrow \t ~~, ~~ \beta_Z \to -\beta_Z )
\Bigg ] + ( 1 \leftrightarrow 2 )  \Bigg ]
~ , \label{Amix1+-+0}
\eqa
\bqa
&& A^{(\tchi_1\tchi_2 2)}_{+-+0}
(\beta_Z,\t,\u; m_1^2, m_2^2)
= -\frac{2(\t-\u+\beta_Z \s)}{\mzd \s} \Bigg \{
2\s C_0^{111}(\s)+\frac{(\s+4 \mzd)Y}{\s_4} D_{ZZ}^{1221}(\t,\u)
\nonumber \\ &&
+E_1^{12}(\s,\u)+E_1^{12}(\s,\t)
-(m_1^2-m_2^2)\s [D_{ZZ}^{1211}(\s,\u)+D_{ZZ}^{1211}(\s,\t)]
+ ( 1 \leftrightarrow 2 )  \Bigg \} ~ , \label{Amix2+-+0}
\eqa
\bqa
&& A^{(\tchi_1\tchi_2 1)}_{++00}
(\beta_Z,\t,\u; m_1^2, m_2^2)
= \frac{32 \mzd}{\s_4} \Bigg \{
\frac{2Y}{\t_1\u_1} -\frac{1}{\s\u_1^2}
[2 \mzd Y +(\u-\t) (\u^2-\mz^4)]B_Z^{12}(\u)
\nonumber \\ &&
-\frac{1}{\s\t_1^2}[2 \mzd Y +(\t-\u) (\t^2-\mz^4)]B_Z^{12}(\t)
-\frac{Y}{\s^2} E_2^{12}(\t, \u) \Bigg \}
+\frac{4}{\s_4 \mzd} \Bigg [  2 m_1^2 [8 \mz^4
\nonumber \\ &&
-2 \s (m_1^2-m_2^2) +\s\s_4]C_0^{111}(\s)
+ 2 \s m_1^2 (m_1^2-m_2^2)^2 \tilde F^{12}(\s,\t,\u)
+\frac{(m_1^2+m_2^2)\s_4 \mzd}{\s} E_2^{12}(\t, \u)
\nonumber \\ &&
+\frac{1}{\s} \Big \{ (m_1^4+m_2^4) \s (\t+\mzd)(\u+\mzd)
+4 (m_1^2+m_2^2)\mz^4 Y
+2m_1^2 m_2^2 \s [Y+\t (\t+\mzd)
\nonumber \\ &&
+\u (\u+\mzd)]
\Big \} D_{ZZ}^{1221}(\t,\u)
+ \Big \{ \frac{2}{\u_1}[8 \mz^6 m_1^2
-\u_1^2 (m_1^2-m_2^2)^2] C_Z^{211}(\u)
\nonumber \\ &&
+ 2 m_1^2 [m_1^2 \mzd (4 \mzd -3 \s)
+m_2^2 (s_2^2+ \s\mzd)
+2 \s_2 \mz^4] D_{ZZ}^{1211}(\s,\u)
\nonumber \\ &&
+( \u \leftrightarrow \t )
\Big \} + ( 1 \leftrightarrow 2 )  \Bigg ]
~, \label{Amix1++00}
\eqa
\bqa
&& A^{(\tchi_1\tchi_2 2)}_{++00}
(\beta_Z,\t,\u; m_1^2, m_2^2)
= -\frac{4}{\mzd} \Bigg [
2 \s C_0^{111}(\s) +\frac{2\mzd}{\s} E_2^{12}(\t,\u)
\nonumber \\ &&
+(m_1^2+m_2^2) \s_2\tilde F^{12}(\s,\t,\u)
-2 \mzd (m_1^2-m_2^2) [D_{ZZ}^{1211}(\s,\u)
+D_{ZZ}^{1211}(\s,\t)]
\nonumber \\ &&
 + ( 1 \leftrightarrow 2 )  \Bigg ]
~, \label{Amix2++00}
\eqa
\bqa
&& A^{(\tchi_1\tchi_2 1)}_{+-+-}
(\beta_Z,\t,\u; m_1^2, m_2^2)
=
-\frac{16 [\s_2 Y+\beta_Z\mzd \s (\u-\t)]}{\s_4\t_1\u_1}
\nonumber \\ &&
+\frac{8 \s_2 [Y-\s(\s_4 +\beta_Z(\u-\t))]}{\s_4Y}
[B_Z^{11}(\s) +B_Z^{22}(\s)]
-\frac{4\s\s_2}{\s_4 Y^2} \Big [
3 \mz^8 +(\t^2+\u^2)(Y-3 \mz^4)
\nonumber \\ &&
+\t^4 +\t^2\u^2 +\u^4 -\beta_Z\s(\t-\u)
(\t^2+\t\u+\u^2 -2 \mz^4)] [C_0^{111}(\s)+C_0^{222}(\s)]
\nonumber \\ &&
+\frac{4\s}{\s_4 Y^2} \Big \{
-\s_4 [(\t^2-\mz^4)^2 + (\u^2-\mz^4)^2 -(\t\u +\mz^4)^2
+\t\u\s_2^2]
\nonumber \\ &&
+\beta_Z (\t-\u)[2Y^2
+(\t^2+\u^2)(4 Y +\t^2 +\u^2 -\t\u)-2 \mz^4 \t\u]\Big \}
[C_{ZZ}^{121}(\s)+C_{ZZ}^{212}(\s)]
\nonumber \\ &&
- \Big \{ 16 \Big [ \Big (
\frac{\mz^4 (\u-\t)}{\s_4\u_1^2}-\frac{1}{2} \Big )
(1-\beta_Z)+\frac{\mzd}{\s_4}
\Big (1- \frac{2\mz^4}{\u_1^2} \Big )
-\frac{2\u\beta_Z}{\s_4} -\frac{\u^2}{Y\s_4}
[\s_4
\nonumber \\ &&
+\beta_Z (\u-\t)] \Big ] B_Z^{12}(\u)
+ \frac{2}{\s_4 Y^2} \Big [
\s_2 (Y^2-2\u^3 \s_2)+4 \mz^4 \u (Y+2\u^2)
\nonumber \\ &&
-\beta_Z\s[Y^2
-2\u^2 (2 \mz^4+\u\s_2)] \Big ]
[E_1^{12}(\s,\u)+E_1^{21}(\s,\u)]
+( \u \leftrightarrow \t ~~, ~~ \beta_Z \to -\beta_Z ) \Big \}
\nonumber \\ &&
+ \frac{4}{\s_4Y^2} \Bigg [
-4Y (m_1^2-m_2^2) [Y-\s (\s_4+\beta_Z (\u-\t))]B_Z^{11}(\s)
\nonumber \\ &&
-\s \Big \{ (2Y-\s\s_4)(m_1^2-m_2^2)^2 [2(m_1^2-m_2^2)
-3\s_2]-(m_1^2-m_2^2)[4 \s_4 Y m_1^2
\nonumber \\ &&
+2\mz^4 (2\t\u +\mz^4)
+(\t^2+\u^2)(4\t\u-10\mz^4)+4Y^2 +3(\t^4
+\u^4)]
+2m_1^2 \s_4\s_2 Y
\nonumber \\ &&
+\beta_Z (\u-\t) \{ \s (m_1^2-m_2^2)^2
[3\s_2-2(m_1^2-m_2^2)]
-(m_1^2-m_2^2) [4 Y m_1^2 +\s (3 (\u^2+\t^2)
\nonumber \\ &&
+4Y)]
+ 2\s_2 Y m_1^2 \} \Big \}C_0^{111}(\s)
-\s \Big \{  \s_4 (m_1^2-m_2^2) [4 Y +3 (\t-\u)^2]
(m_1^2-m_2^2-\s_2)
\nonumber \\ &&
+2 m_1^2 \s_4^2 Y
 +\beta_Z (\u-\t) \{(m_1^2-m_2^2)
[10Y +3 (\t-\u)^2](m_1^2-m_2^2-\s_2)
\nonumber \\ &&
+2 m_1^2\s_4 Y \} \Big \} C_{ZZ}^{121}(\s)
+\frac{1}{\s} \{ 4 m_1^2 m_2^2 \mzd \s Y^2
+2 (m_1^4+m_2^4)\s (3\mzd -\s)Y^2
\nonumber \\ &&
-(m_1^2+m_2^2)\s_2 Y^3
+(m_1^2-m_2^2)^4 \s^2 (2Y -\s\s_4)
+2 (m_1^2-m_2^2)(m_1^4-m_2^4) (Y-\s\s_4) Y\s
\nonumber \\ &&
+\beta_Z\s(\t-\u) (m_1^2-m_2^2)^2
[\s^2 (m_1^2-m_2^2)^2
+2(m_1^2+m_2^2) Y\s +Y^2] \} D_{ZZ}^{1221}(\t,\u)
\nonumber \\ &&
+\Bigg [ \frac{2}{\s\u_1}
\Big \{ \s \u_1^2 (m_1^2-m_2^2)^2
[Y (\u_1-\mzd)+3\u (\u^2-2\mz^4)+3\t \mz^4]
+(m_1^2-m_2^2) \u_1^2 [ 2\mz^8 (\s-\u_1)
\nonumber \\ &&
-2 \mz^6\u(\t+\t_1)
+2\mzd \u^2 (\u-2\mzd)^2 -\s\u^2
( 14 \mz^4 -\s^2+ 2\mzd \s +2 \u\s_4)]
\nonumber \\ &&
+ 2 Y m_1^2 [2 \mz^8 (\mzd -4\u)+\mzd \s_4
(\mz^4\s +\s^2 \u -4 \mzd \u^2) +4\mz^6 \u (2\s-\u)
+\mzd \s^2 \u\u_1
\nonumber \\ &&
+\u^3 (2 \mzd \u +\s^2 -8 \mz^4)]
-\beta_Z \s \{ \u_1^3 (m_1^2-m_2^2)^2 (Y+3\mzd \t_1 +3\u\u_1)
\nonumber \\ &&
+(m_1^2-m_2^2) \s\u\u_1^2 (4\mz^4 -\t\u -3\u^2)
+2m_1^2\s Y [\u\u_1 (\u+\mzd)-\mzd Y]\} \Big \} C_Z^{211}(\u)
\nonumber \\ &&
-\frac{1}{2} \Big \{
-2\s (m_1^2-m_2^2)^4 (2Y-\s\s_4) +4 (m_1^6-m_2^6)\s
(\mz^4 \t-5\mz^4 \u -\s_2 \t\u +2\u^3)
\nonumber \\ &&
-8 Y m_1^6 (Y-\s\s_4) +4 m_1^2 m_2^2 (m_1^2 - m_2^2)
(2\mz^8 -14 \mz^6 \s + 5\mz^4 \s^2- 8 \mz^6 \u
+26 \mz^4\s\u
\nonumber \\ &&
-\s^3\u +12 \mz^4 \u^2 -10 \mzd \s\u^2 +\s^2\u^2-8 \mzd \u^3
-2\s\u^3 +2\u^4)+8 Y m_1^4 m_2^2 (Y-\s\s_4)
\nonumber \\ &&
+6 \s (m_1^2-m_2^2)^2 (\mz^8-6\mz^4 \u^2 +\t^2\u^2-2\u^3\s_2)
+16 Y m_1^4 (-\mzd Y+\mz^4 \t-3 \mz^4 \u -\u^2 \s_2)
\nonumber \\ &&
-8 Y m_1^2 m_2^2 (\mz^4 \t-7 \mz^4 \u +\u\s_2^2-\u^2\s_2)
+(m_1^2-m_2^2) \s [-\mz^8\s_2 -10\u\mz^4 Y-2\mz^4 \t^2\u
\nonumber \\ &&
+\t^3\u^2 -\u^3 (4Y -8\s\s_4+ \t^2)]
+2 Y m_1^2 [-\mz^4 (8 Y + \s_2^2 +16  \u^2) +6\t^2\u^2
+9 \t\u^3 +4 \u^4 +\t^3\u]
\nonumber \\ &&
-\beta_Z\s \{ 2\s (m_1^2-m_2^2)^4 (\t-\u)
+4\s (m_1^4-m_2^4)(m_1^2+m_2^2) (\mz^4+\t\u -2\u^2)
+8 Y (\t-\u)m_1^6
\nonumber \\ &&
-32 m_1^4 m_2^2\u_1 (\u\u_1+\mzd\t_1)
-8m_1^2 m_2^4 (2\mz^6-3\mz^4 \s +6\mzd \u\u_1
+\s^2 \u +3\s\u^2-2u^3)
\nonumber \\ &&
-(m_1^2-m_2^2)[ 12 \s\u (m_1^2-m_2^2) (\u^2-\mz^4)
+16 Y m_1^2  (\u^2-\mz^4) + \s Y^2 +4\s \u^2 (Y
\nonumber \\ &&
+2(\u^2-\mz^4))] -2 m_1^2 Y [\t Y +\u(3Y +4(\u^2-\mz^4))]
\} \Big \} D_{ZZ}^{1211}(\s,\u)
\nonumber \\ &&
+~( \u \leftrightarrow \t ~~, ~~ \beta_Z \to -\beta_Z )
\Bigg ] + ( 1 \leftrightarrow 2 )  \Bigg ]
~, \label{Amix1+-+-}
\eqa
\bqa
&& A^{(\tchi_1\tchi_2 2)}_{+-+-}
(\beta_Z,\t,\u; m_1^2, m_2^2)
= \frac{8}{\s_4} \Big \{ 2Y -\s [\s_4 +\beta_Z (\t-\u)]\Big \}
D_{ZZ}^{1221}(\t, \u)
~. \label{Amix2+-+-}
\eqa

\newpage

\clearpage
\newpage

\end{document}